\documentclass[10pt]{article}

\usepackage[british]{babel}
\usepackage[margin=1in]{geometry}
\usepackage[T1]{fontenc}
\usepackage[utf8]{inputenc}
\usepackage{lmodern}
\usepackage{microtype}
\usepackage{setspace}
\usepackage{amsmath,amssymb,amsthm,mathtools,bm}
\usepackage{booktabs,tabularx,multirow,array}
\usepackage{graphicx}
\usepackage{float}
\usepackage{caption}
\usepackage{subcaption}
\usepackage{enumitem}
\usepackage{hyperref} 
\usepackage{placeins}
\usepackage{natbib}
\usepackage{threeparttable} 
\usepackage{algorithm}

\usepackage{xcolor}
\definecolor{mjdblue}{RGB}{31,119,180}
\definecolor{vgorange}{RGB}{255,127,14}
\definecolor{forestgreen}{RGB}{44,160,44}
\definecolor{brickred}{RGB}{214,39,40}
\definecolor{royalpurple}{RGB}{148,103,189}
\definecolor{chocolate}{RGB}{140,86,75}
\definecolor{rosepink}{RGB}{227,119,194}
\definecolor{mediumgray}{RGB}{127,127,127}
\definecolor{olivegreen}{RGB}{188,189,34}
\definecolor{cyanblue}{RGB}{23,190,207}

\hypersetup{colorlinks=true,linkcolor=blue,citecolor=blue,urlcolor=blue}
\newcommand{\todoins}[1]{\textcolor{red}{\emph{TBD}}}
\newcolumntype{Y}{>{\raggedright\arraybackslash}X}

\AtBeginDocument{%
	\setlength{\abovedisplayskip}{8pt plus 2pt minus 3pt}%
	\setlength{\belowdisplayskip}{8pt plus 2pt minus 3pt}%
	\setlength{\abovedisplayshortskip}{5pt plus 2pt minus 2pt}%
	\setlength{\belowdisplayshortskip}{5pt plus 2pt minus 2pt}%
}

\title{Valuation of GLWB--LTC Annuities with L\'evy Equity Dynamics,\\ Stochastic Interest Rates and Health-State Transitions}

\author{
	Andrea Molent\thanks{Dipartimento di Scienze Economiche e Statistiche,
		Universit\`a degli Studi di Udine, Udine, Italy.\\
		Email: \href{mailto:andrea.molent@uniud.it}{andrea.molent@uniud.it}.
		ORCID: \href{https://orcid.org/0000-0002-0887-826X}{0000-0002-0887-826X}.}
}

\date{}

\begin{document}
\maketitle

\begin{abstract}
This paper develops a valuation framework for guaranteed lifetime withdrawal
benefit (GLWB) contracts with long-term care (LTC) features when the reference
fund follows exponential L\'evy dynamics and the short rate follows the
Hull--White model. The contract combines financial guarantees, longevity
protection, health-contingent LTC payments, and surrender optionality, requiring
the joint treatment of jump risk, stochastic discounting, and disability risk.
The numerical method couples a recombining Hull--White trinomial tree with an implicit--explicit (IMEX) finite difference scheme. The framework
incorporates a seven-state health model, annual fees, LTC payments, guaranteed
withdrawals, and bang-bang policyholder actions, and is benchmarked against
Monte Carlo simulation. Numerical results show that the
hybrid tree--IMEX method delivers stable long-maturity prices consistent with
simulation benchmarks. They also show that L\'evy equity dynamics and stochastic interest rates have
a material impact on fair fees and surrender incentives, and affect the
composition of contract value. The findings highlight the importance of modelling
financial tail risk and interest-rate risk jointly when pricing long-term
insurance guarantees with LTC-contingent benefits.
\end{abstract}



\section{Introduction}

The design of retirement products that jointly address longevity risk, investment risk, and long-term care  needs remains a central challenge in actuarial science. Traditional annuities provide lifetime income, but they typically offer limited protection against severe late-life care expenses. Stand-alone LTC insurance, on the other hand, suffers from persistent demand and supply frictions, including adverse selection, complexity, and high perceived cost. Bundled products that combine retirement income and LTC protection have therefore emerged as a promising line of innovation \citep{murtaugh2001,brownwarshawsky2013}.

Within this line of research, \citet{hsieh2018} introduced the variable life care annuity with guaranteed lifetime withdrawal benefits, a contract that couples guaranteed withdrawals with LTC-contingent payments and values it by Monte Carlo simulation with variance reduction. More recently, \citet{apicella2024} extended the GLWB--LTC setting by allowing dynamic withdrawals and surrender under stochastic interest rates, and proposed a tree-based numerical method. On a parallel track, \citet{goudenege2024} showed how variable annuities with embedded guarantees can be valued in a L\'evy equity market with Hull--White interest rates by combining tree methods and finite-difference techniques for solving the
partial integro-differential equation (PIDE) associated with variable-annuity
pricing. More recently, \citet{chen2026} contributed to the literature on LTC-related
guaranteed withdrawal products under richer contract features and advanced
numerical methods.

Despite these advances, a clear gap remains. The GLWB--LTC literature has not yet been
 fully integrated with a richer financial setting in which the underlying fund is driven by a
L\'evy process and the short rate is stochastic. This gap matters for two reasons. First, jump risk is relevant for long-dated insurance guarantees because downside events and heavy tails materially affect guarantee costs. Second, stochastic interest rates are essential for contracts with long maturities because discounting and intertemporal risk transfer become sensitive to the shape and volatility of the term structure \citep{goudenege2024}.
  
The paper is also connected to the literature on variable annuities and
policyholder behaviour, to which Anna Rita Bacinello made important
contributions, including the unifying valuation framework of
\citet{bacinello2011}. Moreover, in the numerical experiments, we use the
option-implied L\'evy parameter sets adopted by \citet{bacinello2016} for the
valuation of variable-annuity guarantees as financial benchmarks.

Building on this literature, this paper addresses that gap by developing a
risk-neutral valuation framework for GLWB--LTC contracts in a L\'evy--Hull--White
environment. The model combines: 
(i) a seven-state health-transition structure aligned with the benchmark LTC
literature; (ii) an annual contract operator with fees on the account value and
the benefit base, LTC payments, guaranteed withdrawals, surrender, and
bang-bang policyholder actions; and (iii) a hybrid backward solver based on a
recombining interest-rate tree and an IMEX discretisation of the L\'evy
generator. The framework supports several L\'evy specifications and allows the
comparison of static, mixed, dynamic, and full-dynamic withdrawal strategies.

The contribution of the paper is methodological and numerical. First, the paper
formulates the valuation of GLWB--LTC contracts in a financial environment in
which the reference fund follows martingale-corrected exponential L\'evy
dynamics and the short rate follows the Hull--White model. This extends the
GLWB--LTC framework beyond the diffusion and deterministic-rate specifications
typically used in the benchmark literature. Second, the paper adapts the
hybrid tree--IMEX methodology developed for L\'evy variable annuities with
stochastic interest rates to a contract with health-contingent withdrawals,
annual event dates, benefit-base dynamics, surrender optionality, and
multi-state disability transitions. Third, the numerical analysis documents how
the joint presence of calibrated return tails, stochastic discounting, and LTC
contingent benefits affects fair fees, value decomposition, and surrender
incentives.  The contribution is therefore not to introduce a new contractual design, but to
integrate existing GLWB--LTC mechanics with L\'evy--Hull--White valuation in a
reproducible backward numerical framework.

The rest of the paper is organised as follows. Section~\ref{sec:contract}
describes the contract design and the actuarial state variables.
Section~\ref{sec:model} introduces the health, fund, and short-rate dynamics
under the pricing measure. Section~\ref{sec:valuation} formulates the valuation
problem and the annual backward operator. Section~\ref{sec:numerical_method}
presents the hybrid tree--IMEX algorithm and the Monte Carlo benchmark.
Section~\ref{sec:numerics} reports the numerical results. Section~\ref{sec:conclusion}
concludes.

\section{Contract design and actuarial state variables}
\label{sec:contract}

\subsection{Product rationale}

We consider a single-premium GLWB--LTC contract purchased at age $x_0$ with initial premium $P>0$. The product combines a guaranteed lifetime withdrawal benefit with health-contingent LTC payments. The contract is designed so that the policyholder receives a guaranteed annual withdrawal while alive and, in severe disability states, an additional LTC payment linked to the benefit base. This structure is closely related to the life care annuity logic of \citet{hsieh2018} and to the generalised GLWB--LTC framework of \citet{apicella2024}.

Two state variables govern the contract evolution: the account value \(A_t\),
which reflects the value of the investment account, and the benefit base
\(B_t\), which determines the guaranteed payments. By homogeneity, the
valuation problem can be reduced by fixing the benefit base and solving it in
terms of the normalised account ratio; we keep the full notation here for
clarity and discuss the reduction in
Section~\ref{subsec:similarity_reduction}.

\subsection{Health states}
\label{subsec:health_states}

Following the benchmark specification used by \citet{hsieh2018}
and \citet{apicella2024}, the policyholder's health status is described by
a seven-state process \(M_t\), with
\[
M_t \in \{1,2,3,4,5,6,7\}.
\]
The states, their role in the disability model, and the corresponding LTC
eligibility rule are summarised in Table~\ref{tab:health_states}. The LTC
benefit is triggered only in the severe disability states, namely states
\(4\), \(5\), and \(6\), while no LTC payment is made in states \(1\), \(2\),
and \(3\).

This convention is consistent with the multi-state disability structure used
in the GLWB--LTC literature and allows the contract to distinguish between
moderate health deterioration and disability levels that generate long-term
care payments. The live-state process is not restricted to monotone
deterioration: transitions between disability states may also represent
recoveries, consistently with the continuous-time multi-state disability model
of \citet{pritchard2006}. Cognitive impairment is not introduced as a separate
state; it is captured only insofar as it leads to sufficient ADL impairment to
trigger LTC eligibility. The transition dynamics of the health process are
specified in Section~\ref{sec:model}.

\begin{table}[t]
	\centering 
	\begin{tabular}{clcc}
		\toprule
		State & Description & Model role & LTC payment \\
		\midrule
		1 & Healthy & Non-claiming & No \\
		2 & IADL impairment & Non-claiming & No \\
		3 & 1--2 ADL impairments & Non-claiming & No \\
		4 & 3--4 ADL impairments & Claiming & Yes \\
		5 & 5--6 ADL impairments & Claiming & Yes \\
		6 & Institutionalized & Claiming & Yes \\
		7 & Dead & Absorbing & No \\
		\bottomrule
	\end{tabular}
	\caption{\small\label{tab:health_states}Health states and LTC eligibility.}
\end{table}

\subsection{Anniversary mechanics and policyholder actions}
\label{subsec:anniversary_mechanics}
\label{subsec:policyholder_action}
The contract evolves on annual anniversaries. At inception, \(n=0\),
the initial premium $P$ determines the account value and the benefit base;
fees are charged immediately, while no LTC payment or withdrawal is made.
At each subsequent anniversary \(n\geq 1\), conditional on the policyholder
being alive, the annual event sequence is:
\begin{enumerate}[label=\alph*)]
	\item the account value is reduced by fees proportional to the account
	value and to the benefit base, while the benefit base is left unchanged;
	\item if the policyholder is in an LTC-eligible state, the LTC payment
	is made;
	\item the admissible withdrawal or surrender decision is applied.
\end{enumerate}
 
Death is handled through the annual health-state transition. If the policyholder 
dies between two consecutive anniversaries, the contract terminates at the next
 anniversary, when the corresponding death payoff is settled. Hence, in the 
 backward recursion, the death-state contribution is included in the 
 health-mixing step before the financial continuation is propagated backward.

Let \((A_n^-,B_n^-)\) denote the account value and benefit base immediately
before the \(n\)-th anniversary. Fees are deducted at the beginning of the
anniversary event sequence, and we denote by \((A_n^{(1)},B_n^{(1)})\) the
resulting post-fee state. Fees consist of an account-value fee \(\alpha\) and a
benefit-base charge \(\beta\), the latter being proportional to the benefit
base but deducted from the account value. Hence,
\[
A_n^{(1)}
=
\max\left\{(1-\alpha)A_n^- - \beta B_n^-,0\right\},
\qquad
B_n^{(1)}=B_n^- .
\]
The next event is the LTC payment, after which the state is denoted by
\((A_n^{(2)},B_n^{(2)})\). 

The LTC payment depends on the health state and is triggered only when the
policyholder is in one of the LTC-eligible states, namely states \(4,5,\) and
\(6\). The LTC cash flow at anniversary \(n\) is specified as
\[
L_n = c(1+\pi)^n B_n^{(1)}
\mathbf{1}_{\{M_n\in\{4,5,6\}\}},
\]
where \(c\) is the LTC payout rate and \(\pi\) is the indexation parameter. The
post-LTC account value and benefit base are therefore
\begin{equation}\label{eq:LTC_pay}
A_n^{(2)}
=
\max\left\{
A_n^{(1)}-L_n,\,0
\right\},
\qquad
B_n^{(2)}=B_n^{(1)}.
\end{equation}

The cash flow \(L_n\) is paid in full whenever the policyholder is in an
LTC-eligible state. The account value is reduced by the amount of the payment
only up to zero. Hence, if \(L_n>A_n^{(1)}\), the excess
\(L_n-A_n^{(1)}\) is borne by the insurer. This convention is consistent
with the guarantee nature of the rider: the account value finances the payment
as long as it is positive, while the insurer covers the shortfall after account
depletion.

The guaranteed withdrawal at anniversary \(n\) is denoted by \(G_n\). In the
baseline specification,
\[
G_n  = g_n B_n^{(2)},
\]
where \(g_n\) may be either constant or age-dependent. In particular, in the numerical
experiments in Section \ref{sec:numerics},
 we use the constant-rate case, \(g_n\equiv g\), unless
otherwise stated.

After fees and LTC payments, the policyholder may choose among a finite set of
actions. Rather than allowing a continuum of partial withdrawals, we restrict
attention to the three actions
\[
\gamma_n\in\{0,1,2\},
\]
where \(\gamma_n=0\) denotes no withdrawal, \(\gamma_n=1\) denotes the
guaranteed withdrawal, and \(\gamma_n=2\) denotes full surrender. This
discretisation of the withdrawal control is motivated by the bang-bang
structure of GLWB-type contracts. In particular, \citet{bacinello2024} prove
that, under a total-wealth maximization criterion, the optimal GLWB action is
attained at one of three extreme choices: no withdrawal, withdrawal of the
guaranteed amount, or full surrender. This structure is not generic to all
variable annuities, but is consistent with the GLWB--LTC analysis of
\citet{apicella2024}, who find the same pattern numerically. We therefore
adopt this three-action control set in the implementation.

For a given action \(\gamma_n\), let \(Y_n^{(\gamma_n)}\) denote the
action-dependent cash flow paid to the policyholder at anniversary \(n\),
excluding the LTC payment \(L_n\). The latter has already been paid and
incorporated into the post-LTC state \((A_n^{(2)},B_n^{(2)})\) in
Equation~\eqref{eq:LTC_pay}. The post-action account value and benefit base are denoted by
\((A_n^+,B_n^+)\). The three actions operate as follows.

If \(\gamma_n=0\), no cash flow is paid and a roll-up bonus may be credited to
the benefit base:
\[
Y_n^{(0)}=0,
\qquad
A_n^{+}=A_n^{(2)},
\qquad
B_n^{+}=(1+\rho_n)B_n^{(2)},
\]
where \(\rho_n\) denotes the roll-up bonus rate credited to the benefit base
when no withdrawal is taken.

If \(\gamma_n=1\), the policyholder takes the guaranteed withdrawal:
\[
Y_n^{(1)}=G_n,
\qquad
A_n^{+}=\max\{A_n^{(2)}-G_n,0\},
\qquad
B_n^{+}=B_n^{(2)}.
\]

If \(\gamma_n=2\), the policyholder surrenders the contract. In the mixed
strategy used in the main numerical analysis, surrender is evaluated after the
scheduled guaranteed withdrawal. If \(\chi_n\) denotes the surrender penalty
applied to the residual account value, the surrender cash flow is
\[
Y_n^{(2)}
=
G_n
+
(1-\chi_n)\max\{A_n^{(2)}-G_n,0\},
\]
and the contract terminates:
\[
A_n^{+}=0,
\qquad
B_n^{+}=0.
\]

As an illustrative choice for the numerical experiments, we adopt the
surrender-penalty schedule
\[
\chi_n = 0.01\max\{8-n,0\}, \qquad n=0,1,\ldots,T.
\]
We stress that the penalty applies only to the residual account value after
the scheduled guaranteed withdrawal.

Different exercise rules are obtained by restricting the admissible action set \(\Gamma_n\).
The static strategy fixes \(\Gamma_n=\{1\}\), so that the policyholder always
takes the guaranteed withdrawal. The mixed strategy uses
\(\Gamma_n=\{1,2\}\), allowing surrender at anniversary dates after the
guaranteed withdrawal. The dynamic strategy uses
\(\Gamma_n=\{0,1,2\}\), adding the no-withdrawal bonus option. Finally, the full
dynamic strategy also allows surrender during the intra-year financial
propagation. These strategy definitions are used in the numerical section to
separate the value of scheduled withdrawals, anniversary surrender, bonus-driven
deferral, and intra-year surrender.

If the annual health transition between anniversaries \(n-1\) and \(n\) leads
to the absorbing death state, that is \(M_{n-1}\in\{1,\ldots,6\}\) and
\(M_n=7\), the death payoff is paid at anniversary \(n\), and the contract
terminates. We denote the death payoff by \(D_n(A,B)\), explicitly displaying
the dependence of the guaranteed withdrawal \(G_n\) on the benefit base \(B\):
\begin{equation}\label{eq:death_terminal_payoff}
	D_n(A,B)=\max\{A,G_n(B)\}.
\end{equation}
At death, or when the maximum age is reached, the contract pays the larger of
the remaining account value and one scheduled guaranteed withdrawal. This gives
the backward recursion a simple terminal payoff and ensures that a depleted
account does not imply a zero terminal value when the withdrawal guarantee is
still active.

\section{Stochastic framework under the pricing measure}
\label{sec:model}

\subsection{Health dynamics}

The health state process is modelled as a time-inhomogeneous continuous-time Markov chain. Let
\[
P_x(s,t) = \bigl(p_{i,j}^{x}(s,t)\bigr)_{i,j=1}^{7}
\]
be the transition matrix between ages $x+s$ and $x+t$. Following the benchmark setup in \citet{hsieh2018} and \citet{apicella2024}, the process is assumed piecewise time-homogeneous over each policy year, so that the yearly transition matrix is obtained by matrix exponentiation:
\begin{equation} \label{eq:health_transition_matrix}
P_x(n,n+1)=\exp\{Q_x(n)\},\qquad n=0,1,\ldots,T-1, 
\end{equation}
where $Q_x(n)$ is the annual intensity matrix applying between attained
ages $x+n$ and $x+n+1$.
The transition intensities are built from the parametric specification in
\citet{pritchard2006}, consistently with the GLWB--LTC benchmark literature.
In particular, \citet{hsieh2018} adopt this health-state classification and
transition-intensity framework, while \citet{apicella2024} employ the same
underlying transition matrices in their GLWB--LTC tree model. Using this common
biometric basis keeps the present numerical results comparable with these
benchmark studies, while isolating the additional effects generated by L\'evy
equity dynamics and Hull--White stochastic interest rates.

The resulting transition probabilities are used as biometric pricing
assumptions. They should be interpreted as best-estimate actuarial probabilities
under the maintained disability and mortality model, rather than as probabilities
derived from traded financial instruments. Accordingly, the fair fees reported
below are market-consistent with respect to the hedgeable financial risks, while
being conditional on the chosen biometric basis. No additional loading for
non-hedgeable disability, mortality, expenses, capital costs, or profit margins
is included.

Independence between the health process and the financial risk factors is
assumed under the pricing measure used for valuation. This assumption is
standard in the benchmark literature and is maintained here to keep the pricing
problem tractable \citep{hsieh2018,goudenege2024}.

\subsection{L\'evy equity dynamics}
\label{subsec:levy_equity}

Let \(S_t\) denote the reference fund value. We work under a fixed pricing
measure \(\mathbb{Q}\), acknowledging that the presence of jumps and
non-hedgeable biometric risk generally makes the market incomplete. This
modelling convention is standard in exponential L\'evy pricing models and in
actuarial-finance valuation frameworks with prescribed biometric transition
assumptions; see, for example, \citet{cont2004} and
\citet{hsieh2018,apicella2024}. Under \(\mathbb{Q}\), the reference fund is
specified as
\begin{equation}\label{eq:dynamic_S}
S_t
=
S_0
\exp\left\{
\int_0^t (r_u-q)\,du
+
X_t
-
K_X(1)t
\right\},
\end{equation}
where \(q\geq 0\) is the dividend yield, \(X=\{X_t\}_{t\geq 0}\) is a
L\'evy process, and \(K_X\) denotes its cumulant-generating exponent per unit
time:
\[
\mathbb{E}^{\mathbb{Q}}\!\left[e^{zX_t}\right]
=
\exp\{tK_X(z)\},
\]
for all \(z\) in the domain where the expectation is finite. 
The term \(-K_X(1)t\) in Equation~\eqref{eq:dynamic_S}  is the exponential martingale correction. It ensures
that the dividend-adjusted discounted fund value
\[
e^{-\int_0^t (r_u-q)\,du}S_t
=
S_0 e^{X_t-K_X(1)t}
\]
is a martingale under \(\mathbb{Q}\).    

The framework can accommodate several exponential L\'evy specifications, provided
that the corresponding L\'evy triplet and exponential compensator \(K_X(1)\) are
well defined.  These include, for example, Merton jump--diffusion, Variance Gamma,
NIG, and CGMY dynamics. 

Let \(\mathcal{L}_X\) denote the infinitesimal generator of the L\'evy component.
For a sufficiently regular test function \(u\), it can be written in the form
\[
\mathcal{L}_X u(y)
=
\frac{\sigma^2}{2}u_{yy}(y)
+
\mu u_y(y)
+
\int_{\mathbb{R}\setminus\{0\}}
\left[
u(y+z)-u(y)-z\mathbf{1}_{\{|z|\leq 1\}}u_y(y)
\right]\nu(dz),
\]
where \((\mu,\sigma,\nu)\) is the L\'evy triplet under the chosen truncation
convention. Conditional on a short-rate value \(r\), the generator acting on the
log-account variable is therefore
\begin{equation}\label{eq:Gr}
\mathcal{G}_r u(y)
=
(r-q-K_X(1))u_y(y)
+
\mathcal{L}_X u(y).
\end{equation}
This is the operator discretised in the IMEX step of the hybrid numerical
scheme.

\subsection{Hull--White short rate}
\label{subsec:hull_white}

The short rate follows the one-factor Hull--White model
\begin{equation}\label{eq:sde_HW}
dr_t
=
\kappa_{\mathrm{HW}}
\bigl(\theta(t)-r_t\bigr)\,dt
+
\omega_{\mathrm{HW}}\,dW_t^r,
\qquad
r_0>0,
\end{equation}
where \(\kappa_{\mathrm{HW}}>0\) is the mean-reversion speed,
\(\omega_{\mathrm{HW}}>0\) is the short-rate volatility, and \(\theta(t)\) is
chosen to fit the initial zero-coupon term structure.  

The Hull--White Brownian motion and the L\'evy equity component are assumed to
be independent. This factor-separation assumption is consistent with the hybrid
tree--IMEX approach used for L\'evy variable-annuity valuation with stochastic
interest rates in \citet{goudenege2024}, where the account-value dimension is
propagated conditionally on the interest-rate state. The assumption is not meant
to rule out empirical dependence between equity and interest-rate risks; rather,
it defines a tractable benchmark in which the effects of equity jump risk and
stochastic discounting can be isolated.

\section{Valuation problem and annual backward recursion}
\label{sec:valuation}

\subsection{Value function}
\label{subsec:value_function}

Let \(V_n(a,b,m,r)\) denote the pre-anniversary contract value at anniversary
\(n\), conditional on
\[
A_n^-=a,\qquad B_n^-=b,\qquad M_n=m\in\{1,\ldots,6\},
\qquad r_n=r.
\]
Throughout this section, lower-case variables such as \(a\), \(b\), \(m\), and \(r\)
denote generic arguments of value functions and operators, while upper-case
symbols such as \(A_n\) and \(B_n\) denote the stochastic contract state
variables. When annual events are described, superscripts distinguish the
timing within the anniversary, as in \(a^{(1)}\), \(a^{(2)}\), and
\(a_n^{+,\gamma}\).

The valuation problem is implemented through a backward recursion that combines
health-state mixing, financial propagation, and the annual contract operator.
Between anniversaries, the account value and the short rate are propagated under
the L\'evy--Hull--White financial dynamics. At anniversary dates, the
continuation value is transformed by the contract operator described in
Section~\ref{subsec:anniversary_mechanics}, which applies fees, LTC payments,
withdrawals, surrender, and, when admissible, the no-withdrawal bonus option.

The backward recursion starts from the terminal value at the maximum contract
horizon. For each live health state \(m\in\{1,\ldots,6\}\), we set
\[
V_T(a,b,m,r)=D_T(a,b),
\]
where \(D_T\) is the death or terminal payoff defined in
Equation~\eqref{eq:death_terminal_payoff}. In the backward recursion, policy year \(n\) denotes the interval between
anniversaries \(n\) and \(n+1\). Then, for
\(n=T-1,T-2,\ldots,0\), the backward step over policy year \(n\) maps the
next-anniversary value function \(V_{n+1}\) into the current-anniversary value
function \(V_n\). This step is split into three parts.

First, future values are averaged over the possible health transitions during
the year. We denote this health-mixing step by
\[
\widehat V_{n+1}=\mathcal H_n V_{n+1}.
\]
Second, the health-mixed value is propagated from anniversary \(n+1\) back to
anniversary \(n\) under the financial dynamics of the account value and the
short rate, including discounting. We denote this financial propagation by
\[
\widetilde V_n=\mathcal F_n \widehat V_{n+1}.
\]
Finally, the annual contract events at anniversary \(n\) are applied through
the annual operator \(\mathcal A_n\). For \(n\geq 1\), this operator includes
fees, LTC payments, withdrawals, surrender, and any admissible policyholder
action; at \(n=0\), it reduces to the inception fee step. Hence, the backward
recursion is
\[
V_n=\mathcal A_n \widetilde V_n,
\qquad n=T-1,T-2,\ldots,0.
\]

\subsection{Health mixing and financial continuation}
\label{subsec:health_mixing}

The health process is updated at annual dates. Therefore, before propagating
the contract value backward through the financial dynamics over policy year
\(n\), the future value \(V_{n+1}\) is first averaged over the possible health
states at anniversary \(n+1\). This health-mixing step is performed pointwise
in the financial variables.

For a policyholder entering the contract at age \(x_0\), the health transition
over policy year \(n\) is described by
\[
p^{x_0}_{i,j}(n,n+1)
=
\mathbb P(M_{n+1}=j\mid M_n=i),
\qquad i,j\in\{1,\ldots,7\}.
\]
Equivalently, this is the one-year transition from attained age \(x_0+n\) to
attained age \(x_0+n+1\). The yearly transition matrix \(P_{x_0}(n,n+1)\) then updates the health state
from \(n\) to \(n+1\).

Fix a live health state \(m\in\{1,\ldots,6\}\) at anniversary \(n\). We denote
by \(\widehat V_{n+1}(a,b,m,r)\) the value at anniversary \(n+1\) after mixing
over the health transitions that can occur during year \(n\), conditional on
being in state \(m\) at the beginning of the year. It is given by
\begin{equation}
	\label{eq:hV}
	\widehat V_{n+1}(a,b,m,r)
	=
	\sum_{m'=1}^6
	p^{x_0}_{m,m'}(n,n+1)
	V_{n+1}(a,b,m',r)
	+
	p^{x_0}_{m,7}(n,n+1)
	D_{n+1}(a,b).
\end{equation}
The summation accounts for transitions to live health states, while the last
term accounts for death during the year. The index \(m\) in
\(\widehat V_{n+1}\) therefore denotes the live health state at the beginning
of policy year \(n\); the possible health states at anniversary \(n+1\) have
already been averaged out in \eqref{eq:hV}.

The health-mixed value \(\widehat V_{n+1}\) is then propagated backward from
anniversary \(n+1\) to anniversary \(n\) under the financial dynamics of the
account value and the short rate, including discounting. The resulting function
\(\widetilde V_n\) gives the discounted financial continuation value as a
function of the post-anniversary state \((A_n^+,B_n^+)\). It is the quantity
evaluated inside the annual contract operator. Equivalently,
\[
\widetilde V_n=\mathcal F_n\widehat V_{n+1},
\]
where \(\mathcal F_n\) denotes the financial propagation operator over policy
year \(n\). In probabilistic form,
\begin{equation}
	\label{eq:tV}
	\widetilde V_n(a,b,m,r)
	=
	\mathbb{E}^{\mathbb{Q}}
	\left[
	\exp\left\{-\int_n^{n+1} r_u\,du\right\}
	\widehat V_{n+1}
	\left(
	A_{n+1}^{-},b,m,r_{n+1}
	\right)
	\,\middle|\,
	A_n^{+}=a,\ B_n^{+}=b,\ M_n=m,\ r_n=r
	\right].
\end{equation}
Here \(b\) remains fixed between anniversaries, while the account value and the
short rate evolve under the financial model. Thus, \(\widehat V_{n+1}\) captures
the annual health-state mixing, whereas \(\widetilde V_n\) is the discounted
financial continuation value entering the annual operator. The PIDE
characterisation in the next subsection provides the corresponding analytic
description of the same financial propagation step.

\subsection{PIDE characterisation of the financial continuation value}
\label{subsec:pide_characterisation}

The financial propagation step can also be characterised through a partial
integro-differential equation, which provides a continuous-time representation
of the operator \(\mathcal F_n\) over policy year \(n\).

Fix a live health state \(m\in\{1,\ldots,6\}\) and a benefit base \(b\). Since
both are fixed between anniversaries, for \(t\in[n,n+1]\) we define the
intra-year financial continuation value as
\[
\mathcal C_n(t,y,r;m,b),
\qquad y=\log a,
\]
where \(a>0\) is the account value. After solving the PIDE backward from
\(n+1\) to \(n\), the financial continuation value entering the annual operator
is obtained by evaluating \(\mathcal C_n\) at the current anniversary.

For the static, mixed, and dynamic strategies, no surrender is allowed during
the intra-year financial propagation. Hence, on the open interval
\((n,n+1)\), \(\mathcal{C}_{n}\) satisfies the backward PIDE
\begin{equation}
	\label{eq:PIDE}
	\frac{\partial \mathcal{C}_{n}}{\partial t}
	+
	\mathcal G_r \mathcal{C}_{n}
	+
	\kappa_{\mathrm{HW}}\bigl(\theta(t)-r\bigr)
	\frac{\partial \mathcal{C}_{n}}{\partial r}
	+
	\frac{\omega_{\mathrm{HW}}^2}{2}
	\frac{\partial^2 \mathcal{C}_{n}}{\partial r^2}
	-
	r\mathcal{C}_{n}
	=0,
\end{equation}
with terminal condition
\[
\mathcal{C}_{n}(n+1,y,r;m,b)
=
\widehat V_{n+1}(e^y,b,m,r).
\]
Here \(\mathcal G_r\) is the log-account operator defined in
Equation~\eqref{eq:Gr}, acting on the \(y\)-variable, and
\(\widehat V_{n+1}\) is the health-mixed value introduced in
Equation~\eqref{eq:hV}. The discounted financial continuation value entering
the annual operator is therefore
\[
\widetilde V_n(a,b,m,r)
=
\mathcal{C}_{n}(n,\log a,r;m,b).
\]
 
   The PIDE is posed on the natural positive account domain
\(a\in(0,\infty)\). The truncation of this domain, the explicit treatment of
the depleted-account state, and the artificial boundary conditions used in the
discrete scheme are described in Section~\ref{subsec:conditional_pide_imex}.

For the full dynamic strategy, the annual operator is the same as in the
dynamic strategy, but surrender is also allowed during the intra-year financial
propagation. Let
\[
S^{\mathrm{intra}}_n(a)=(1-\chi_n)a
\]
be the intra-year surrender payoff during policy year \(n\). This payoff does
not include the scheduled guaranteed withdrawal, since guaranteed withdrawals
are paid only at anniversary dates. The full-dynamic intra-year continuation value
\(\mathcal C_n^{\mathrm{full}}\) is then characterised, on
\((n,n+1)\), by the obstacle problem
\begin{equation}\label{eq:obstacle}
\max\left\{
\frac{\partial \mathcal C_n^{\mathrm{full}}}{\partial t}
+
\mathcal G_r\mathcal C_n^{\mathrm{full}}
+
\kappa_{\mathrm{HW}}\bigl(\theta(t)-r\bigr)
\frac{\partial \mathcal C_n^{\mathrm{full}}}{\partial r}
+
\frac{\omega_{\mathrm{HW}}^2}{2}
\frac{\partial^2 \mathcal C_n^{\mathrm{full}}}{\partial r^2}
-
r\mathcal C_n^{\mathrm{full}},
\,
S^{\mathrm{intra}}_n(e^y)
-
\mathcal C_n^{\mathrm{full}}(t,y,r;m,b)
\right\}
=0,
\end{equation}
with terminal condition
\[
\mathcal C_n^{\mathrm{full}}(n+1,y,r;m,b)
=
\widehat V_{n+1}(e^y,b,m,r).
\]
In the continuation region,
\[
\mathcal{C}_{n}^{\mathrm{full}}(t,y,r;m,b)
>
S^{\mathrm{intra}}_n(e^y),
\]
the linear PIDE holds. In the stopping region, the value coincides with the
intra-year surrender payoff. Thus, the full dynamic strategy differs from the
dynamic strategy only through the continuation value entering the annual
operator:
\[
\widetilde V_n(a,b,m,r)
=
\mathcal{C}_{n}^{\mathrm{full}}(n,\log a,r;m,b)
\]
under full dynamic exercise.

\subsection{Annual contract operator}
\label{subsec:annual_operator}

The annual operator \(\mathcal{A}_n\) maps the financial continuation value
\(\widetilde V_n\) into the pre-anniversary value \(V_n\). Starting from the
generic pre-anniversary state \((a,b,m,r)\), the fee step gives
\[
a^{(1)}
=
\max\{(1-\alpha)a-\beta b,0\},
\qquad
b^{(1)}=b.
\] 
The LTC cash flow is
\[
L_n(a,b,m)
=
c(1+\pi)^n b^{(1)}
\mathbf 1_{\{m\in\{4,5,6\}\}},
\]
and the post-LTC state is
\[
a^{(2)}
=
\max\{a^{(1)}-L_n(a,b,m),0\},
\qquad
b^{(2)}=b^{(1)}.
\]
The guaranteed withdrawal amount at this stage is determined by the post-LTC
benefit base:
\[
G_n(b^{(2)})=g_n b^{(2)}.
\]

After fees and LTC payments, the policyholder chooses an action
\(\gamma\in\Gamma_n\). For a given action, let
\[
\left(a_n^{+,\gamma},b_n^{+,\gamma}\right)
\]
be the post-action state, and let \(Y_n^{(\gamma)}\) be the cash flow generated
by the action itself. The three possible actions are specified as follows. For no withdrawal,
\[
Y_n^{(0)}=0,
\qquad
a_n^{+,0}=a^{(2)},
\qquad
b_n^{+,0}=(1+\rho_n)b^{(2)}.
\]
For the guaranteed withdrawal,
\[
Y_n^{(1)}=G_n(b^{(2)}),
\qquad
a_n^{+,1}=\max\{a^{(2)}-G_n(b^{(2)}),0\},
\qquad
b_n^{+,1}=b^{(2)}.
\]
For full surrender,
\[
Y_n^{(2)}
=
G_n(b^{(2)})+(1-\chi_n)\max\{a^{(2)}-G_n(b^{(2)}),0\},
\qquad
a_n^{+,2}=0,
\qquad
b_n^{+,2}=0.
\]

The annual operator is therefore
\[
V_n(a,b,m,r)
=
\max_{\gamma\in\Gamma_n}
\left\{
L_n(a,b,m)
+
Y_n^{(\gamma)}
+
\mathbf 1_{\{\gamma\neq 2\}}
\widetilde V_n
\left(
a_n^{+,\gamma},
b_n^{+,\gamma},
m,
r
\right)
\right\}.
\]
The indicator \(\mathbf 1_{\{\gamma\neq 2\}}\) makes explicit that, under full
surrender, the contract terminates and no continuation value is added.
The maximisation is taken over the admissible action set \(\Gamma_n\) associated
with the selected exercise strategy, as defined in Section~\ref{subsec:anniversary_mechanics}.
The full dynamic strategy uses the same anniversary action set as the dynamic
strategy. The difference is that the financial continuation value
\(\widetilde V_n\) is obtained from the obstacle problem in
\eqref{eq:obstacle}, rather than from the linear PIDE, because surrender
is also allowed during the intra-year propagation. This
distinction separates the value of annual exercise rights from the additional
value generated by intra-year surrender opportunities.

At inception, \(n=0\), the initial premium determines the account value and the
benefit base, and the implementation applies only the fee step. Thus, no LTC
payment, withdrawal, or surrender decision is applied at \(n=0\), and the
inception operator reduces to
\[
V_0(a,b,m,r)
=
\widetilde V_0
\left(
\max\{(1-\alpha)a-\beta b,0\},
b,
m,
r
\right).
\]
\subsection{Similarity reduction}
\label{subsec:similarity_reduction}
The contract is homogeneous in the account value and benefit base. This type of
similarity reduction is standard in the PDE literature on GLWB valuation; see,
for instance, \citet{forsythVetzal2014}, who exploits the homogeneity of the contract
value with respect to the investment account and the guarantee account in order
to reduce the dimensionality of the pricing problem. 
For any \(\eta>0\),
\[
V_n(\eta a,\eta b,m,r)=\eta V_n(a,b,m,r).
\]
The same homogeneity applies to the annual cash flows and post-action mappings,
because fees, guaranteed withdrawals, LTC payments, surrender payments, and
death or terminal payoffs are all proportional to either the account value or
the benefit base.
This property allows the monetary dimension to be reduced by fixing the benefit
base and solving the problem in terms of the normalised account-to-base ratio.
Equivalently, one may write
\[
V_n(a,b,m,r)
=
b\,v_n\left(\frac{a}{b},m,r\right),
\]
for a reduced value function \(v_n\). The numerical implementation exploits
this reduction by solving on a one-dimensional account grid, combined with the
health states and the Hull--White rate nodes. This reduction is essential for
making the hybrid tree--IMEX recursion feasible over the long maturities
considered in the numerical section.

\section{Hybrid tree--IMEX method and Monte Carlo benchmark}
\label{sec:numerical_method}

The numerical method builds on the hybrid tree--IMEX  approach of
\citet{goudenege2024}, which combines a recombining tree for the
Hull--White interest-rate factor with finite-difference/IMEX propagation in the
account-value dimension, in the spirit of the hybrid tree/finite-difference
framework of \citet{briani2019}. The method is adapted here to the GLWB--LTC
setting by incorporating health-state mixing, annual contract events,
benefit-base dynamics, LTC-contingent payments, and policyholder surrender
optionality.

The time axis is divided into yearly intervals, each interval is further
subdivided into \(N_t\) substeps, and the solver alternates two components:
\begin{enumerate}[label=\alph*)]
	\item a recombining tree approximation for the Hull--White short rate;
	\item a conditional IMEX discretisation of the one-dimensional PIDE in the
	log-account variable.
\end{enumerate}
At each year boundary, the algorithm applies health-state mixing and the
backward anniversary operator. 
 
 \subsection{Hull--White tree}
 \label{subsec:hw_tree}
 
 The short-rate factor is approximated by a recombining trinomial tree. The use
 of tree approximations for stochastic-rate variable annuity valuation is in
 line with previous contributions on GMWB/GLWB products with Hull--White
 interest rates, including \citet{molent2020} and \citet{goudenege2021}. This
 tree provides the conditional rate dynamics used in the financial propagation
 step.
 
 At each substep $\ell$, the tree provides a finite set of rate nodes
 \begin{equation}\label{eq:r_grid}
 \{r_{\ell,k}: k=1,\ldots,N_r(\ell)\},
 \end{equation}
 together with successor indices and transition probabilities. We denote the
 successor nodes from $(\ell,k)$ by
 \[
 k_u(\ell,k), \qquad k_m(\ell,k), \qquad k_d(\ell,k),
 \]
 and the corresponding probabilities by
 \[
 p_u(\ell,k), \qquad p_m(\ell,k), \qquad p_d(\ell,k).
 \]
 
 The transition probabilities are chosen under the pricing measure by matching
 the first two conditional moments of the Hull--White short rate over one
 substep. More precisely, at node $(\ell,k)$, let
 \[
 \mu^r_{\ell,k}
 =
 \mathbb E^{\mathbb Q}
 \left[
 r_{\ell+1}\mid r_\ell=r_{\ell,k}
 \right],
 \qquad
 \left(\sigma^r_\ell\right)^2
 =
 \operatorname{Var}^{\mathbb Q}
 \left[
 r_{\ell+1}\mid r_\ell=r_{\ell,k}
 \right]
 \]
 denote the conditional mean and variance implied by the Hull--White dynamics.
 The successor indices $k_u(\ell,k)$, $k_m(\ell,k)$, and $k_d(\ell,k)$ are
 selected on the recombining lattice around the conditional mean, and the
 probabilities are determined so that
\[
\left\{
\begin{aligned}
	p_u(\ell,k)+p_m(\ell,k)+p_d(\ell,k)
	&=1,\\
	p_u(\ell,k)r_{\ell+1,k_u}
	+
	p_m(\ell,k)r_{\ell+1,k_m}
	+
	p_d(\ell,k)r_{\ell+1,k_d}
	&=\mu^r_{\ell,k},\\
	p_u(\ell,k)
	\left(r_{\ell+1,k_u}-\mu^r_{\ell,k}\right)^2
	+
	p_m(\ell,k)
	\left(r_{\ell+1,k_m}-\mu^r_{\ell,k}\right)^2
	+
	p_d(\ell,k)
	\left(r_{\ell+1,k_d}-\mu^r_{\ell,k}\right)^2
	&=\left(\sigma^r_\ell\right)^2 .
\end{aligned}
\right.
\]
 This moment-matching construction gives the risk-neutral trinomial transition
 probabilities used in the tree.
 
In the backward recursion, the Hull--White tree is used to compute the
conditional expectation over successor rate nodes.
Let \(W_{\ell,k}\) denote a generic vector of grid values on the account grid
at substep \(\ell\) and Hull--White node \(k\).
 Then, for any collection
$\{W_{\ell+1,k'}\}_{k'}$ of successor-node value vectors, we define
\begin{equation}\label{eq:operator_E}
	\mathcal E^{HW}_{\ell,k} W_{\ell+1}
	=
	p_u(\ell,k) W_{\ell+1,k_u(\ell,k)}
	+
	p_m(\ell,k) W_{\ell+1,k_m(\ell,k)}
	+
	p_d(\ell,k) W_{\ell+1,k_d(\ell,k)} .
\end{equation}
 This operator represents the non-discounted conditional expectation over the
 successor rate nodes. The local short rate $r_{\ell,k}$ then enters the
 account-value PIDE \eqref{eq:PIDE} through the drift coefficient and enters discounting
 through the scalar factor $\exp(-r_{\ell,k}\Delta t)$. In the implementation,
 this discount factor is applied once per substep, after the conditional IMEX
 update in the account-value dimension. This ordering represents the nodewise
 discounting over the substep and makes explicit that no double discounting is
 performed.
 
 The recombining structure keeps the number of rate nodes manageable over long
 maturities. In the numerical implementation, the future-node indices,
 transition probabilities, short-rate values, and discount factors are
 precomputed and stored before the backward induction starts. This caching step
 avoids recomputing the Hull--White tree quantities during the repeated IMEX
 updates.

\subsection{Conditional PIDE and IMEX discretisation}
\label{subsec:conditional_pide_imex}

We now focus on the IMEX discretisation of the PIDE
\eqref{eq:PIDE} in the account-value dimension. Following the similarity reduction in
Section~\ref{subsec:similarity_reduction}, the natural state variable is the
normalised account-to-base ratio. In the implementation, the reduced problem is
stored in the equivalent monetary coordinate obtained by fixing the reference
benefit base at \(P\). Thus, writing \(x=a/P\), using \(\log a\) or \(\log x\)
leads to the same finite-difference operator, since
\[
\log a=\log x+\log P.
\]
In the notation below, \(b\) is kept as a parameter to preserve the connection
with the unreduced formulation. In the actual grid implementation, however, the
reference benefit base is fixed at \(b=P\), and values for other benefit-base
levels are recovered by homogeneity.

The strictly positive part of the account grid is built in logarithmic
coordinates, and the depleted-account state is added as a separate explicit
node:
\begin{equation}\label{eq:a_grid}
a_0=0,\qquad a_j=\exp\{y_{\min}+(j-1)\Delta y\},\qquad j=1,\ldots,N_y.
\end{equation}
Equivalently, since the reference benefit base is fixed at \(P\), the associated
normalised grid is \(x_j=a_j/P\).  The
finite-difference/IMEX step is applied only to the strictly positive nodes
\(a_j>0\), \(j=1,\ldots,N_y\). The node \(a_0=0\) is not a logarithmic grid
point and is propagated separately.

The boundary treatment is as follows. At the lower boundary, the
depleted-account state \(a=0\) is represented by a separate continuation value
\[
\mathcal{C}_{n}^0(t,r;m,b).
\]
This value should not be interpreted as \(\mathcal{C}_{n}(t,0,r;m,b)\), since
\(y=0\) corresponds to account value \(a=1\), not to the depleted-account
state. Rather, it represents the limiting value associated with account
depletion:
\[
\lim_{a\downarrow 0}
\mathcal{C}_{n}(t,\log a,r;m,b)
=
\mathcal{C}_{n}^0(t,r;m,b).
\]
The terminal condition at the depleted-account node is
\[
\mathcal{C}_{n}^0(n+1,r;m,b)
=
\widehat V_{n+1}(0,b,m,r).
\]

From an implementation standpoint, we denote by \(W_{\ell,k}\) the vector of
continuation values on the positive account grid at substep \(\ell\) and
Hull--White node \(k\). Its \(j\)-th component approximates
\(\mathcal C_n(t_\ell,\log a_j,r_{\ell,k};m,b)\), for
\(j=1,\ldots,N_y\). The separate value at the depleted-account node is denoted
by \(W^0_{\ell,k}\).
In the discrete tree--IMEX scheme, the zero-account value is propagated as
\[
W^0_{\ell,k}
=
e^{-r_{\ell,k}\Delta t}
\sum_{s\in\{\mathrm u,\mathrm m,\mathrm d\}}
p_s(\ell,k)
W^0_{\ell+1,k_s(\ell,k)}.
\]
Thus, once the account is depleted, it remains zero under the fund dynamics,
while the remaining contract value is still propagated through stochastic
discounting and future contractual guarantees.  

At the upper boundary, the contract value is assumed to be asymptotically
linear. Following the standard finite-difference treatment of artificial
boundaries for PIDE problems in exponential L\'evy models \citep{cont2005},
we impose a Neumann-type linear boundary condition on the truncated
computational domain. Since the stored grid is uniform in the log-account
coordinate, this condition is implemented through the linear closure
\[
W_{N_y,\ell,k}
=
2W_{N_y-1,\ell,k}
-
W_{N_y-2,\ell,k}.
\] 

To construct the IMEX step, at each Hull--White node \((\ell,k)\) we split the
conditional log-account generator into a local drift-diffusion part and a
nonlocal jump part:
\[
\mathcal G_{r_{\ell,k}}u
=
\mathcal D_{r_{\ell,k}}u+\mathcal Ju.
\]
Using the L\'evy triplet \((\mu,\sigma,\nu)\) and the compensated generator
introduced in Section~\ref{subsec:levy_equity}, the two components are
\[
\begin{aligned}
	\mathcal D_{r_{\ell,k}}u(y)
	&=
	\frac{\sigma^2}{2}u_{yy}(y)
	+
	\left(r_{\ell,k}-q-K_X(1)+\mu\right)u_y(y),\\
	\mathcal Ju(y)
	&=
	\int_{\mathbb R\setminus\{0\}}
	\left[
	u(y+z)-u(y)-z\mathbf 1_{\{|z|\leq 1\}}u_y(y)
	\right]\nu(dz).
\end{aligned}
\]

At each substep, the value is first averaged over the successor Hull--White
nodes using the operator \(\mathcal E^{HW}_{\ell,k}\) introduced in
\eqref{eq:operator_E}:
\[
\overline W_{\ell,k}
=
\mathcal E^{HW}_{\ell,k} W_{\ell+1}.
\]
Then, conditional on the current short-rate node \(r_{\ell,k}\), one IMEX step
is applied in the account-value dimension:
\[
\left(I-\Delta t\,\mathcal D^h_{r_{\ell,k}}\right)
\widetilde W_{\ell,k}
=
\left(I+\Delta t\,\mathcal J^h\right)
\overline W_{\ell,k}.
\]
Here \(\mathcal D^h_{r_{\ell,k}}\) denotes the finite-difference
discretisation of the local drift-diffusion operator
\(\mathcal D_{r_{\ell,k}}\), while \(\mathcal J^h\) denotes the discrete
nonlocal jump operator. The local part is treated implicitly, since
\(\widetilde W_{\ell,k}\) appears on the left-hand side through
\(\mathcal D^h_{r_{\ell,k}}\). The jump part is treated explicitly, since it is
evaluated on the already known value \(\overline W_{\ell,k}\) on the
right-hand side. The discrete jump operator \(\mathcal J^h\) uses interpolation on the account
grid. Values required below the first positive node are linked to the explicit
zero-account continuation value \(W^0_{\ell,k}\), while values required above
the upper log-grid are tied to the upper boundary value.
Finally, the substep discount factor is applied:
\[
W_{\ell,k}
=
e^{-r_{\ell,k}\Delta t}\widetilde W_{\ell,k}.
\]
Thus, the short-rate node \(r_{\ell,k}\) enters both the local drift and the
discount factor, but discounting is applied only once per substep.

The scheme follows the IMEX logic of \citet{cont2005}: the local part is treated
implicitly, while the jump integral is treated explicitly. This avoids dense
linear systems from the nonlocal operator; the resulting systems are banded and
are solved independently for each health state and Hull--White node.

After annual contract events, the post-event account value is evaluated on the
account grid by interpolation. Values between the zero node and the first
positive node are interpolated linearly; inside the positive grid, interpolation
is performed in the account value after locating the neighbouring log-grid
points.

This conditional propagation is repeated backward within each policy year,
starting from the health-mixed continuation value at the next anniversary.
Once the current anniversary is reached, the annual contract operator $\mathcal{A}_n$ is applied as described in 
Section~\ref{subsec:annual_operator}.

\subsection{Health mixing and annual step}
\label{subsec:health_mixing_annual_step}

The yearly backward step combines the three components introduced above:
health-state mixing, financial propagation, and the annual contract operator.
Starting from \(V_{n+1}\), the solver first constructs the health-mixed
continuation value using the yearly health-transition matrix. This value is
then propagated backward within the policy year through the Hull--White tree
and the conditional IMEX step in the account-value dimension. Finally, at the
current anniversary, the annual contract operator is applied separately for
each live health state and each rate node, with the inception step at \(n=0\)
restricted to the initial fee deduction.

This decomposition keeps the recursion modular while preserving the ordering of
events in the contract. The complete backward recursion is summarised in
Algorithm~\ref{alg:backward_imex_tree}.

\begin{algorithm}[H]
	\centering
	\fbox{
		\begin{minipage}{0.92\textwidth} 
			
			For \(n=T-1,\ldots,0\):
			\begin{enumerate}[label=\alph*)]
				\item Construct the health-mixed continuation value from \(V_{n+1}\)
				using the yearly transition matrix \(P_{x_0}(n,n+1)\), including
				transitions to the absorbing death state.
				
				\item Propagate the continuation value backward through the substeps
				of policy year \(n\), for each live health state and rate node. At
				each substep, compute the non-discounted Hull--White conditional
				expectation over successor rate nodes, apply one conditional IMEX
				step in the account-value dimension, and then apply the substep
				discount factor once.
				
				\item At the current anniversary, apply the annual contract operator. For
				\(n\geq 1\), this includes fees, LTC payments, and the admissible policyholder
				action set \(\Gamma_n\). For \(n=0\), only the inception fee step is applied.
				The maximising action is selected using the total contract value whenever an
				action is admissible.
				
				\item Store the resulting value \(V_n\) for all live health states,
				account-grid points, and Hull--White rate nodes.
			\end{enumerate}
		\end{minipage}
	}
	\caption{Backward tree--IMEX recursion.}
	\label{alg:backward_imex_tree}
\end{algorithm}

\subsection{Monte Carlo benchmark}
\label{subsec:mc_benchmark} 

As an independent benchmark, we also implement a Monte Carlo simulation under
the pricing measure.   Along each path, the
Hull--White short-rate factor is simulated using its Gaussian transition law,
while the L\'evy equity component is generated from the increment distribution
of the selected model; see, for example, \citet{brigo2006} for Hull--White
simulation and \citet{cont2004} for exponential L\'evy models. The annual
health-state transitions are simulated from the yearly transition matrices
\(P_{x_0}(n,n+1)\).

Given the simulated financial and health trajectories, the contract cash-flow
rules are applied path by path. At each anniversary, the account value and the
benefit base are updated according to the fee, LTC-payment, withdrawal, and
terminal-payoff rules described in Section~\ref{subsec:anniversary_mechanics}.
Discounted pathwise cash flows are then averaged to obtain the Monte Carlo
price estimate, and confidence intervals are computed from the corresponding
sample standard error.

In the present implementation, the Monte Carlo benchmark is used for the static
strategy, where the policyholder action is prescribed pathwise. Strategies with
surrender or other dynamic controls would require an additional approximation
of continuation values, for instance through least-squares Monte Carlo. The
static benchmark is therefore used as a direct validation of the financial
simulation, health transitions, discounting, and contractual cash-flow
mechanics.

\section{Numerical results}
\label{sec:numerics}

This section presents the numerical results obtained with the hybrid
tree--IMEX method. The analysis has four objectives. First, we document the
convergence of the numerical scheme and benchmark it against Monte Carlo
simulation in the static case. Second, we quantify the marginal value generated
by the LTC rider and decompose the contract value into its main cash-flow
components. Third, we measure the impact of replacing simpler financial
benchmarks with L\'evy equity dynamics and Hull--White stochastic rates. 
Finally, we examine how jump risk affects both fair fees and surrender
incentives under the mixed strategy.
 
\subsection{Baseline specification}
\label{subsec:baseline_specification}

The financial analysis focuses on four equity specifications: Geometric
Brownian Motion (GBM), Merton jump--diffusion (MJD), Variance Gamma (VG), and
CGMY. The corresponding parameters are listed in
Table~\ref{tab:levyparams}. They are taken from \citet{bacinello2016},
 who calibrate these model
specifications to the same cross-section of S\&P 500 option prices.
Here, these parameter sets are used as an illustrative market-calibrated
benchmark for comparing the numerical impact of alternative fund dynamics.
 The purpose of this choice is not to provide a
new market calibration, but to use empirically grounded parameter values that
are comparable across L\'evy specifications and suitable for numerical
comparison.

The Hull--White parameters define an illustrative stochastic-rate benchmark,
consistent with the variable-annuity literature using Hull--White-type interest
rate dynamics; see, for example, \citet{molent2020}, \citet{goudenege2021},
and \citet{goudenege2024}. Throughout the numerical analysis, we use a flat
initial term structure for transparency and numerical comparability, with
\(r_0=2\%\). This choice is made for convenience only and is not a restriction
of the tree--IMEX framework, which can accommodate a general initial
zero-coupon curve through the time-dependent drift in the Hull--White model.
The remaining parameters, \(\kappa_{\mathrm{HW}}=0.20\) and
\(\omega_{\mathrm{HW}}=0.03\), provide a baseline with mean reversion and
non-negligible interest-rate volatility. The impact of these Hull--White
parameters is examined in the sensitivity analysis of
Section~\ref{subsec:financial_effects}.

The health-transition matrices are obtained from the disability model in
\citet{pritchard2006} and are used consistently with the GLWB--LTC literature.
They are available up to attained age $122$, which therefore determines 
the maximum age considered in the numerical recursion.

We consider a benchmark policyholder entering the contract at age \(x_0=60\)
with initial premium \(P=100\). The policyholder starts in the healthy state,
\(M_0=1\), unless otherwise stated. The maximum maturity is set equal to
\(T=122-x_0\), so that the recursion covers the long horizon relevant for
lifetime withdrawal and LTC benefits. The guaranteed withdrawal rate is set to
\(g=0.02\). The LTC payout rate is varied over
\[
c\in\{0,0.03,0.06\},
\]
covering the no-LTC benchmark and two increasing levels of LTC benefits. The
Hull--White and contract parameters are reported in Table~\ref{tab:hwparams}.

The account-value fee \(\alpha\) is treated differently depending on the
numerical experiment. In the pricing and convergence tests reported in
Tables~\ref{tab:convergence_static_mc} and
\ref{tab:convergence_dynamic_strategies}, we set \(\alpha=0\). In the fair-fee
experiments in Table~\ref{tab:fair_alpha_strategies} and in the two-way
financial decomposition of Table~\ref{tab:two_way_financial_decomposition},
\(\alpha\) is calibrated model by model as the value satisfying
\(V_0(\alpha)=P\). In the marginal LTC value and value-decomposition exercises
reported in Figure~\ref{fig:marginal_ltc_value} and
Table~\ref{tab:value_decomposition_mixed}, the no-LTC fair fee is first
calibrated and then kept fixed as the LTC payout rate \(c\) is increased.
 
\begin{table}[H]
	\centering 
	\begin{tabular}{ll}
		\toprule
		Model$\qquad$  & Parameters \\
		\midrule
		
		GBM
		& $\sigma_{\mathrm{GBM}}=0.1361$ \\
		
		MJD
		& $\sigma_{\mathrm{MJD}}=0.1114,\ 
		\lambda_{\mathrm{MJD}}=0.5282,\ 
		\mu^{J}_{\mathrm{MJD}}=-0.1825,\ 
		\sigma^{J}_{\mathrm{MJD}}=0.1094$ \\
		
		VG
		& $\kappa_{\mathrm{VG}}=0.1753,\ 
		\theta_{\mathrm{VG}}=-0.3150,\ 
		\sigma_{\mathrm{VG}}=0.1301$ \\
		
		CGMY 
		& $C_{\mathrm{CGMY}}=0.6817, 
		G_{\mathrm{CGMY}}=18.0293,  
		M_{\mathrm{CGMY}}=57.6250, 
		Y_{\mathrm{CGMY}}=0.8$ \\
		
		\bottomrule
	\end{tabular}
		\caption{\small\label{tab:levyparams} L\'evy model parameters used in the numerical experiments. }
	\vspace{0.5em} 
\end{table}

\begin{table}[H]
	\centering 
\begin{tabular}{lll@{\hspace{1.2cm}}lll}
	\toprule
	Parameter & Symbol & Value 
	& Parameter & Symbol & Value \\
	\midrule
	
	Initial short rate 
	& $r_0$ 
	& $0.02$
	& Entry age 
	& $x_0$ 
	& $60$ \\
	
	Mean reversion 
	& $\kappa_{\mathrm{HW}}$ 
	& $0.20$
	& Premium / benefit base 
	& $P$ 
	& $100$ \\
	
	Short-rate volatility 
	& $\omega_{\mathrm{HW}}$ 
	& $0.03$
	& Guar. withdrawal rate
	& $g$ 
	& $0.02$ \\
	
	Dividend yield 
	& $q$ 
	& $0.00$
	& LTC payout rate 
	& $c$ 
	& $0,\ 0.03,\ 0.06$ \\
	
	Fee on account value 
	& $\alpha$ 
	& $0,\alpha^{\mathrm{fair}}$ 
	& Indexation 
	& $\pi$ 
	& $0.02$
\\
	
	Fee on benefit base 
	& $\beta$ 
	& $0.002$
	& Roll-up bonus 
	& $\rho$ 
	& $g+0.005$ \\
	
	& 
	& 
	& Surrender penalty 
	& $\chi_n$ 
	& $0.01\max\{8\!-\!n,0\}$ \\
	
	\bottomrule
\end{tabular}
	\vspace{0.5em}
\caption{\small \label{tab:hwparams}Hull--White and contract benchmark
	parameters.}
 
\end{table}

Table~\ref{tab:bacinello_hw_rn_moments} reports the annual moments of the log-return distribution obtained by combining the Lévy equity models calibrated by \citet{bacinello2016} with Hull--White stochastic interest rates. The mean includes the martingale correction of the exponential Lévy component, while the Hull--White short-rate contribution is computed under a flat initial yield curve.

The computation of the log-return moments in the presence of stochastic interest rates is not immediate, since the Hull--White short-rate component affects both the mean and the variance of the log-return. The details of the calculation are reported in Appendix~\ref{app:hw_moment_calculation}. 

Table~\ref{tab:bacinello_hw_rn_moments} 
 highlights the different tail profiles generated by the calibrated
models. Relative to the Gaussian benchmark, the MJD and VG specifications
produce higher annual volatility, negative skewness, and excess kurtosis, while
the CGMY specification lies closer to the GBM benchmark but still introduces
moderate left skewness and heavier tails.

\begin{table}[htbp]
	\centering 
	\begin{tabular}{lrrrr}
		\toprule
		Model & Mean & Std. dev. & Skewness & Kurtosis  \\
		\midrule
		GBM--HW  & 0.0109 & 0.1370 &  0.0000 & 3.0000 \\
		MJD--HW  & 0.0030 & 0.1913 & -0.9535 & 4.5514  \\
		VG--HW   & 0.0037 & 0.1860 & -0.7348 & 3.9101 \\
		CGMY--HW & 0.0082 & 0.1567 & -0.3106 & 3.2686  \\
		\bottomrule
	\end{tabular}
	\caption{\small\label{tab:bacinello_hw_rn_moments}Annual log-return moments under martingale-corrected Lévy equity dynamics with Hull--White stochastic interest rates.}
\end{table}

\subsection{Numerical validation and convergence}
\label{subsec:numerical_validation}

The first set of experiments validates the hybrid tree--IMEX method and studies
its convergence as the numerical grid is refined. Table~\ref{tab:numerical_setups} reports the numerical configurations used in
the convergence tests. For the tree--IMEX method, \(\Delta y\) is the spacing
of the logarithmic account grid in Equation \eqref{eq:a_grid}, and \(N_t\) is the
number of time substeps per policy year. The quantity \(N_r^{\max}\) denotes
the maximum number of Hull--White rate nodes in the trinomial tree, whose rate
grid is defined in Equation \eqref{eq:r_grid}. For the Monte Carlo benchmark,
\(N_{\mathrm{MC}}\) denotes the number of simulated paths, while the batch size
is the number of paths processed at a time.
The configurations
A--D progressively refine the log-account grid, increase the number of time
substeps per year, and enlarge the Hull--White lattice. The benchmark
configuration uses a finer grid and is used as the reference tree--IMEX value.
The Monte Carlo benchmark is used as an
independent validation of the static-strategy prices.

The runtimes reported in Table~\ref{tab:numerical_setups} should be interpreted
as target computational budgets used to define the numerical configurations,
rather than as reproducible hardware-independent timings. They were chosen with
reference to preliminary runs performed on a 64-bit HP ProBook 460 16 inch G11
Notebook PC equipped with an Intel(R) Core(TM) Ultra 5 125H processor and
32 GB of RAM at 5600 MT/s.

\begin{table}[H]
	\centering 
	\small
	\begin{tabular}{lllccccc}
		\toprule
		Method & Configuration & Target runtime & $\Delta y$ & $N_t$ & $N_r^{\max}$ & $N_{\mathrm{MC}}$ & Batch size \\
		\midrule
		tree--IMEX 	& A         &$6s$& $0.012$ & $8$   & $43$  & -- & -- \\
					& B         &$25s$& $0.008$ & $12$  & $63$  & -- & -- \\
					& C         &$100s$& $0.004$ & $25$  & $127$ & -- & -- \\
					& D         &$400s$& $0.002$ & $35$  & $177$ & -- & -- \\
					& Benchmark & -- & $0.001$ & $100$ & $503$ & -- & -- \\
					\midrule
	     Monte Carlo & Benchmark & -- & -- & $1$ & -- & $10^6$ & $10^5$\\
		\bottomrule
	\end{tabular} 
	\caption{\small\label{tab:numerical_setups}Numerical configurations for the tree--IMEX method and the Monte Carlo benchmark.}
\end{table}

We first validate the hybrid tree--IMEX method under the static strategy with
\(\alpha=0\). Table~\ref{tab:convergence_static_mc} reports the convergence of
the tree--IMEX method across configurations A--D and the benchmark grid, together
with an independent Monte Carlo estimate and the corresponding 95\% confidence
interval. The comparison is performed for all financial specifications and LTC
payout rates.

\begin{table}[H]
	\centering 
	\small
	\setlength{\tabcolsep}{4pt}\renewcommand{\arraystretch}{0.9}
	\begin{tabular*}{\textwidth}{@{\extracolsep{\fill}}lcccccc}
	\toprule
	\multirow{2}{*}{Configuration}
	& \multicolumn{2}{c}{$c=0$}
	& \multicolumn{2}{c}{$c=0.03$}
	& \multicolumn{2}{c}{$c=0.06$} \\
	\cmidrule(lr){2-3}\cmidrule(lr){4-5}\cmidrule(lr){6-7}
	& Price & Delta & Price & Delta & Price & Delta \\
	\midrule
	
	\multicolumn{7}{l}{\textbf{GBM--HW}} \\
	A     & $99.78$& $0.9399$& $102.25$& $0.9095$& $105.54$& $0.8788$\\
	B     & $99.70$& $0.9394$& $102.16$& $0.9092$& $105.45$& $0.8786$\\
	C     & $99.60$& $0.9388$& $102.07$& $0.9088$& $105.36$& $0.8784$\\
	D     & $99.58$& $0.9388$& $102.05$& $0.9087$& $105.34$& $0.8790$\\
	\addlinespace[6pt]
  \textit{BM }    & $99.54$& $0.9390$& $102.01$& $0.9089$& $105.30$& $0.8793$\\
\textit{MC BM}    & $99.55$& $0.9393$& $102.01$& $0.9092$& $105.31 $& $0.8793$\\
 				  & $\pm0.03$& $\pm0.0005$& $\pm0.03$& $\pm0.0005$& $\pm0.041$& $\pm0.0005$\\
	\midrule
	
		\multicolumn{7}{l}{\textbf{MJD--HW}} \\
	A     & $101.35$& $0.9318$& $104.18$& $0.9057$& $107.71$& $0.8801$\\
	B     & $101.25$& $0.9312$& $104.08$& $0.9052$& $107.62$& $0.8797$\\
	C     & $101.16$& $0.9306$& $103.99$& $0.9048$& $107.52$& $0.8794$\\
	D     & $101.13$& $0.9306$& $103.96$& $0.9048$& $107.50$& $0.8795$\\
	\addlinespace[6pt]
	\textit{BM }   & $101.11$ & $0.9311$ & $103.94$ & $0.9053$& $107.48$& $0.8800$ \\
	\textit{MC BM} & $ 101.08$& $0.9307$& $ 103.91 $&  $0.9050$& $107.46 $& $0.8797  $\\
	& $ \pm 0.05$& $ \pm 0.0006$& $ \pm 0.05 $& $ \pm 0.0006 $& $ \pm0.05$& $ \pm0.0007$\\ 
	\midrule
	
	\multicolumn{7}{l}{\textbf{VG--HW}} \\
	A     & $100.98$& $0.9285$& $103.80$& $0.9019$& $107.34$& $0.8760$\\
	B     & $100.95$& $0.9291$& $103.77$& $0.9026$& $107.30$& $0.8768$\\
	C     & $100.92$& $0.9296$& $103.74$& $0.9033$& $107.26$& $0.8776$\\
	D     & $100.95$& $0.9303$& $103.76$& $0.9041$& $107.28$& $0.8784$\\
	\addlinespace[6pt]
  \textit{BM }   & $100.95$ & $0.9312$ & $103.76$ & $0.9051$& $107.28$& $0.8794$ \\
\textit{MC BM} & $100.95$& $0.9310$& $103.76$&  $0.9050$& $107.29$& $0.8793$\\
& $ \pm0.05 $& $ \pm0.0006 $& $ \pm0.05 $& $ \pm0.0006$& $ \pm0.05$& $ \pm 0.0006$\\ 
  \midrule
  
  	\multicolumn{7}{l}{\textbf{CGMY--HW}} \\
  A     & $100.29$& $0.9390$& $102.98$& $0.9105$& $106.39$& $0.8821$\\
  B     & $100.20$& $0.9370$& $102.83$& $0.9085$& $106.24$& $0.8802$\\
  C     & $100.11$& $0.9361$& $102.76$& $0.9076$& $106.16$& $0.8793$\\
  D     & $100.12$& $0.9359$& $102.74$& $0.9075$& $106.14$& $0.8792$\\
  \addlinespace[6pt]
  \textit{BM }   & $100.09 $ & $0.9358$ & $102.70$ & $0.9074$& $106.09$& $0.8790$ \\
  \textit{MC BM} & $100.09  $& $0.9359  $& $102.71   $&  $0.9074 $& $106.10 $& $0.8792  $\\
  & $ \pm 0.04$& $ \pm 0.0005 $& $ \pm0.04 $& $ \pm 0.0005$& $ \pm0.04$& $ \pm 0.0006$\\ 
	\bottomrule
\end{tabular*}
		\caption{\small\label{tab:convergence_static_mc}Convergence of the tree--IMEX method and comparison with the Monte
		Carlo benchmark under the static strategy. For this test, we use $\alpha=0$. BM denotes the finest tree--IMEX benchmark configuration, and MC BM denotes the Monte Carlo benchmark. Delta denotes the sensitivity of the contract value with respect to the initial account value. The rows labelled \(\pm\) report the Monte Carlo $95\%$ confidence-interval
		half-widths for price and delta.}
	\vspace{0.5em}
	\parbox{0.95\textwidth}{
		\footnotesize
		
	}
	\normalsize
\end{table}

Across all models and LTC payout rates, the tree--IMEX prices stabilise as the
grid is refined, and the configuration-D values are already close to the
benchmark-grid values relative to the initial premium. The same pattern is
observed for the reported deltas, which remain stable across refinements.
Moreover, the benchmark tree--IMEX prices are consistent with the Monte Carlo
95\% confidence intervals. This agreement provides an independent validation of
the financial propagation, the health-state mixing, and the annual contract
operator in the static case.

The convergence analysis is then extended to strategies with policyholder
optionality. Table~\ref{tab:convergence_dynamic_strategies} reports values under
the mixed, dynamic, and full dynamic strategies, again with \(\alpha=0\). The
mixed strategy adds anniversary surrender to the static withdrawal rule. The
dynamic strategy further includes the no-withdrawal bonus option, while the full
dynamic strategy also allows surrender during the intra-year financial
propagation.

\begin{table}[H]
	\centering
	\small
	\setlength{\tabcolsep}{3.5pt}
	\begin{tabular}{lcccc cccc ccc}
		\toprule
		\multirow{2}{*}{Configuration}
		& \multicolumn{3}{c}{$c=0$}&
		& \multicolumn{3}{c}{$c=0.03$}&
		& \multicolumn{3}{c}{$c=0.06$} \\
		\cmidrule(lr){2-4}\cmidrule(lr){6-8}\cmidrule(lr){10-12}
		& {\footnotesize Mixed} & {\footnotesize Dynamic} & {\footnotesize Full dyn.}&\hspace{5mm}
		& {\footnotesize Mixed} & {\footnotesize Dynamic} & {\footnotesize Full dyn.}&\hspace{5mm}
		& {\footnotesize Mixed} & {\footnotesize Dynamic} & {\footnotesize Full dyn.} \\
		\midrule
		
		\multicolumn{10}{l}{\textbf{GBM--HW}} \\
		A &$100.29$&$100.42$&$100.47$&&$102.44$&$103.02$&$103.06$&&$105.61$&$107.19$&$107.21$ \\
		B &$100.24$&$100.37$&$100.42$&&$102.39$&$102.95$&$103.00$&&$105.54$&$107.10$&$107.12$ \\
		C &$100.20$&$100.33$&$100.38$&&$102.34$&$102.90$&$102.94$&&$105.48$&$107.02$&$107.04$ \\
		D &$100.19$&$100.32$&$100.37$&&$102.32$&$102.88$&$102.93$&&$105.46$&$107.00$&$107.02$ \\
		\addlinespace[4pt]
\textit{BM} &$100.18$&$100.30$&$100.35$&&$102.30$&$102.86$&$102.90$&&$105.43$&$106.96$&$106.99$ \\
		 
		 \midrule
		 		\multicolumn{10}{l}{\textbf{MJD--HW}} \\
		 A     & $101.65$ & $101.93$ & $101.98$ && $104.27$ & $105.13$ & $105.16$ && $107.74$ & $109.68$ & $109.69$ \\
		 B     & $101.61$ & $101.88$ & $101.93$ && $104.21$  & $105.05$ & $105.09$  &&  $107.66$ & $109.57$ & $109.58$ \\
		 C     & $101.56$ & $101.82$ & $101.88$ && $104.15$  &$104.98$ & $105.02$ && $107.59$ & $109.48$  & $109.49$ \\
		 D     & $101.55$ & $101.81$ & $101.87$ && $104.14$  & $104.96$ & $105.00$  && $107.57$ & $109.44$ & $109.47$ \\ 
		 \addlinespace[4pt]
		 \textit{BM} & $101.53$ & $101.79$ & $101.85$ && $104.12$ & $104.94$ & $104.98$ && $107.54$ & $109.42$ & $109.46$ \\
		
		\midrule
		
		\multicolumn{10}{l}{\textbf{VG--HW}} \\
		A     & $101.40$ & $101.65$ & $101.70$ && $103.97$ & $104.77$ & $104.81$ && $107.41$ & $109.27$ & $109.28$ \\
		B     & $101.39$ & $101.64$ & $101.69$ && $103.96$ & $104.75$ & $104.79$ && $107.38$ & $109.23$ & $109.25$\\
		C     & $101.38$ & $101.63$ & $101.68$ && $103.94$ & $104.73$ & $104.77$ && $107.35$ & $109.19$ & $109.21$ \\
		D     & $101.40$ & $101.64$ & $101.70$ && $103.95$ & $104.75$ & $104.79$ && $107.37$ & $109.21$ & $109.23$ \\
		\addlinespace[4pt]
		 \textit{BM} & $101.39$ & $101.63$ & $101.69$ && $103.94$ & $104.74$ & $104.78$ && $107.35$ & $109.19$ & $109.21$ \\
		
				\midrule
		
		\multicolumn{10}{l}{\textbf{CGMY--HW}} \\
		A     & $100.73$ & $100.91$ & $100.96$ && $103.08$ & $103.75$ & $103.79$ && $106.37$ & $108.08$ & $108.09$ \\
		B     & $100.69$ & $100.86$ & $100.91$ && $103.02$ & $103.68$ & $103.72$ && $106.30$ & $107.98$ & $108.00$ \\
		C     & $100.65$ & $100.82$ & $100.87$ && $102.97$ & $103.62$ & $103.66$ && $106.23$ & $107.89$ & $107.91$ \\
		D     & $100.64$ & $100.81$ & $100.86$ && $102.96$ & $103.60$ & $103.65$ && $106.21$ & $107.87$ & $107.89$ \\
		\addlinespace[4pt]
		\textit{BM} & $100.63$ & $100.79$ & $100.84$ && $102.94$ & $103.58$ & $103.62$ && $106.19$ & $107.83$ & $107.86$ \\
		
		\bottomrule
	\end{tabular}
\caption{\small \label{tab:convergence_dynamic_strategies}
	Convergence of the proposed tree--IMEX method under the mixed, dynamic, and
	full dynamic strategies. Reported entries are contract values computed with
	\(\alpha=0\). BM denotes the finest tree--IMEX benchmark configuration.}
	\normalsize
\end{table}

The results show that the same refinement pattern is preserved when
policyholder optionality is introduced. Values increase when the admissible
action set is enlarged, as expected, but the numerical convergence remains
stable across configurations. This suggests that the annual maximization step
and the intra-year surrender feature can be incorporated without compromising
the robustness of the tree--IMEX recursion.

Table~\ref{tab:fair_alpha_strategies} reports the fair annual fee
\(\alpha^{\mathrm{fair}}\), calibrated from the condition \(V_0(\alpha)=P\),
under the static, mixed, dynamic, and full dynamic strategies. The comparison is
performed for all financial specifications and LTC payout rates, using both
configuration D and the benchmark grid.

\begin{table}[H]
	\centering
	\small
	\setlength{\tabcolsep}{2pt}
	\resizebox{\textwidth}{!}{%
		\begin{tabular}{lcccc@{\hspace{5mm}}cccc@{\hspace{5mm}}cccc}
			\toprule
			\multirow{2}{*}{Model}
			& \multicolumn{4}{c}{$c=0$}
			& \multicolumn{4}{c}{$c=0.03$}
			& \multicolumn{4}{c}{$c=0.06$} \\
			\cmidrule(lr){2-5}\cmidrule(lr){6-9}\cmidrule(lr){10-13}
			& {\footnotesize Static}
			& {\footnotesize Mixed} 
			& {\footnotesize Dynamic} 
			& {\footnotesize Full dyn.}
			& {\footnotesize Static}
			& {\footnotesize Mixed} 
			& {\footnotesize Dynamic} 
			& {\footnotesize Full dyn.}
			& {\footnotesize Static}
			& {\footnotesize Mixed} 
			& {\footnotesize Dynamic} 
			& {\footnotesize Full dyn.} \\
			\midrule
			
			\textbf{GBM--HW} &&&&&&&&&&&&\\
		   D& $-2.74$ & $2.16$  & $3.35$ & $3.93$
			& $14.32$ & $25.01$  & $28.15$ & $29.23$
	     	& $40.59$ & $58.65$  & $65.54$ & $67.34$ \\
			
 \textit{BM}& $-2.97$ & $1.98$  & $3.16$ & $3.74$
			& $14.06$ & $24.81$  & $27.95$ & $29.03$
			& $40.33$ & $58.44$  & $65.31$ & $67.16$ \\
			
			\midrule
			
						\textbf{MJD--HW} &&&&&&&&&&&&\\
			D& $7.48$ & $17.38$  & $19.01$ & $19.97$
			& $27.97$ & $45.43$ & $49.14$ & $50.82$
			& $57.23$ & $83.80$ & $91.24$ & $93.90$ \\
			
			\textit{BM} & $7.22$ & $17.19 $  & $18.81$ & $19.77$
			& $27.71$ & $45.23 $ &  $48.92$ & $50.64$
			& $56.95$ & $83.58 $ & $91.01$ & $93.71$ \\
			
			\midrule
			
			\textbf{VG--HW}  &&&&&&&&&&&&\\
		   D& $6.27$ & $15.72$ & $17.29$ & $18.19$
			& $26.54$ & $43.46$ & $47.13$ & $48.72$ 
			& $55.27$ & $81.62$ & $89.03$ & $91.58$ \\
			
 \textit{BM} & $6.14$  & $15.62$ & $17.01$& $18.11 $ 
             & $26.20$ & $43.36 $ &  $47.02$ & $48.63$
		     & $55.47$ & $81.49 $ & $88.89$ & $91.47$ \\
			
			\midrule
			
			\textbf{CGMY--HW}  &&&&&&&&&&&&\\
		D	& $0.65$ & $7.21$ & $8.56$ & $9.25$
			& $19.06$ & $32.11$ & $35.47$ & $36.74$ 
			& $46.53$ & $67.68$ & $74.79$ & $76.86$ \\
			
 \textit{BM} & $0.44$  & $7.05$   & $8.39$  & $9.08$ 
      & $18.82$ & $31.91 $ &  $35.26$ & $36.54$ 
& $46.26$ & $67.44$ & $74.52$ & $76.62$ \\
			
			\bottomrule
		\end{tabular}%
	}
	
\caption{\small \label{tab:fair_alpha_strategies}
	Fair annual fee \(\alpha^{\mathrm{fair}}\) under the static, mixed, dynamic,
	and full dynamic strategies, reported
	in basis points.}
	\normalsize
\end{table}
  
 The results show that fair fees increase with the LTC payout rate \(c\) and
  with the degree of policyholder optionality. 
 The move from the static strategy to
 the mixed strategy captures the value of anniversary surrender, while the
 dynamic and full dynamic strategies add the value of bonus-driven deferral and
 intra-year surrender opportunities.
 
 The negative fair fee obtained under the GBM--HW static benchmark with \(c=0\)
 should not be interpreted as an arbitrage or modelling inconsistency. It simply
 indicates that, in this simplified setting, the benefit-base charge \(\beta\)
 is sufficient to more than offset the value of the scheduled guarantee when the
 account-value fee is set to zero. However, this conclusion is not robust to
 richer equity dynamics: under the calibrated L\'evy specifications, the same
 contract requires a positive fair account-value fee. This provides a useful
 warning that a simple diffusion benchmark may understate the cost of long-term
 guarantees and lead to misleading conclusions about fee adequacy.
 
The benchmark and configuration-D values are close, indicating that fair-fee
calibration inherits the stability observed in the direct pricing experiments.
This agreement makes configuration D a practical compromise between accuracy
and computational cost. Therefore, unless otherwise stated, the remaining
numerical experiments are performed using configuration D.

For the actuarial and financial analyses that follow, the mixed strategy is
used as the main reference case. This strategy includes anniversary surrender
optionality while remaining sufficiently parsimonious to isolate the effects of
LTC coverage, jump risk, and stochastic interest rates.

\subsection{Actuarial implications of LTC coverage}
\label{subsec:actuarial_implications_ltc}

\paragraph{Marginal value of LTC coverage}
\label{subsubsec:marginal_value_ltc}

We first quantify the marginal value generated by the LTC rider under the mixed
withdrawal strategy. This strategy includes anniversary surrender while
excluding bonus-driven no-withdrawal decisions and intra-year surrender, thereby
providing a parsimonious setting in which to isolate the incremental value of
LTC coverage.

The analysis is performed under the four financial specifications used
throughout the numerical section: GBM--HW, MJD--HW, VG--HW, and CGMY--HW. The
purpose is not to rank the models, but to check that the actuarial contribution
of the LTC rider is not driven by a single return specification. For each
model, we first calibrate the no-LTC account-value fee and then keep it fixed
as the LTC payout rate is increased.
More precisely, for each Hull--White financial specification
\[
m \in \{\mathrm{GBM\mbox{--}HW},\mathrm{MJD\mbox{--}HW},
\mathrm{VG\mbox{--}HW},\mathrm{CGMY\mbox{--}HW}\},
\]
let \(\alpha_m^{\mathrm{mix,fair}}(c=0)\) denote the fair account-value fee in the
no-LTC benchmark, where the argument of \(\alpha_m^{\mathrm{mix,fair}}\) refers
to the LTC payout rate $c$, namely
\[
V^{\mathrm{mix}}_{0,m}
\left(0;\alpha_m^{\mathrm{mix,fair}}(c=0)\right)=P.
\]
Using the configuration-D no-LTC mixed fair fees reported in
Table~\ref{tab:fair_alpha_strategies}, we have
\begin{equation}
	\label{eq:no_ltc_mixed_fair_fees}
	\begin{aligned}
		\alpha_{\mathrm{GBM\text{--}HW}}^{\mathrm{mix,fair}}(c=0)
		&= \phantom{1}2.16\ \text{bps},
		&
		\alpha_{\mathrm{MJD\text{--}HW}}^{\mathrm{mix,fair}}(c=0)
		&= 17.38\ \text{bps},\\
		\alpha_{\mathrm{VG\text{--}HW}}^{\mathrm{mix,fair}}(c=0)
		&= 15.72\ \text{bps},
		&
		\alpha_{\mathrm{CGMY\text{--}HW}}^{\mathrm{mix,fair}}(c=0)
		&= \phantom{1}7.21\ \text{bps}.
	\end{aligned}
\end{equation}
We then keep \(\alpha_m^{\mathrm{mix,fair}}(c=0)\) fixed and compute the contract value for increasing LTC payout rates \(c\). 
The marginal value of LTC coverage is defined as
\begin{equation}\label{eq:marginal_LTC}
\Delta V_{0,m}^{\mathrm{LTC,mix}}(c)
=
V_{0,m}^{\mathrm{mix}}\left(c;\alpha_m^{\mathrm{mix,fair}}(c=0)\right)-P.
\end{equation}
This quantity measures the additional economic value generated by the health-contingent payments, without re-calibrating the contract after the LTC rider is introduced.

\begin{figure}[H]
	\centering
 	\includegraphics[width=0.7\textwidth]{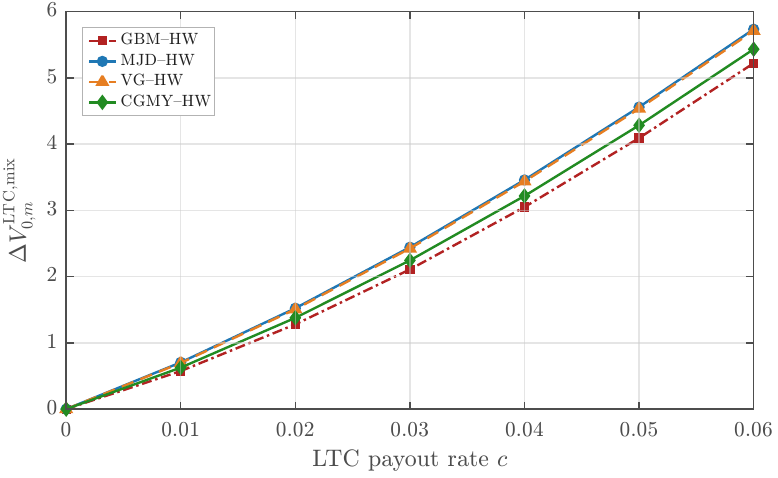}
	
	\caption{\small	\label{fig:marginal_ltc_value}Marginal value of LTC coverage under the mixed withdrawal strategy. For each financial specification, the account-value fee is calibrated in the no-LTC benchmark and then kept fixed as the LTC payout rate \(c\) varies.	
	}
\end{figure}

Figure~\ref{fig:marginal_ltc_value} reports the normalised marginal LTC value over a fine grid of payout rates \(c\). The figure shows that the marginal value of LTC coverage increases with the payout rate. However, this marginal value should not be interpreted as the discounted value of direct LTC payments alone. This point is quantified in the component-decomposition analysis reported later in Table~\ref{tab:value_decomposition_mixed}, where the cases \(c=0.03\) and \(c=0.06\) are decomposed into their main cash-flow components.

\paragraph{Value decomposition}
\label{subsubsec:value_decomposition_ltc}

We next decompose the marginal value of LTC coverage into direct and indirect
components. The purpose of this exercise is to show that the value generated by the LTC
rider is not simply the discounted value of LTC payments. Since LTC benefits are paid
out of the account value, they may also affect the timing of account depletion, the value
of guaranteed withdrawals, the surrender option, and the terminal or death-related payoff.

The decomposition is performed under the mixed withdrawal strategy and uses the
same no-LTC fair-fee convention as the marginal-value analysis: the
account-value fee is kept fixed at the configuration-D values in
Equation~\eqref{eq:no_ltc_mixed_fair_fees}, while the LTC payout rate \(c\) is
increased. For \(c>0\), the marginal LTC value
\(\Delta V_{0,m}^{\mathrm{LTC,mix}}(c)\) is defined as in
Equation~\eqref{eq:marginal_LTC}.

To simplify notation, the dependence on the fixed fee
\(\alpha_m^{\mathrm{mix,fair}}(c=0)\) is suppressed in what follows.
To identify the channels through which this marginal value arises, we write the contract
value as the sum of four discounted cash-flow components:
\[
V_{0,m}^{\mathrm{mix}}(c)
=
V_{0,m}^{\mathrm{GW}}(c)
+
V_{0,m}^{\mathrm{LTC}}(c)
+
V_{0,m}^{\mathrm{SUR}}(c)
+
V_{0,m}^{\mathrm{DB}}(c),
\]
where \(V_{0,m}^{\mathrm{GW}}\) is the value of guaranteed withdrawals,
\(V_{0,m}^{\mathrm{LTC}}\) is the value of direct LTC payments,
\(V_{0,m}^{\mathrm{SUR}}\) is the value of surrender payments, and
\(V_{0,m}^{\mathrm{DB}}\) collects death and terminal payments.

The reported decomposition is expressed in marginal form relative to the no-LTC
benchmark:
\begin{equation}\label{eq:dec_delta}
	\Delta V_{0,m}^{\mathrm{LTC,mix}}(c)
	=
	\Delta V_{0,m}^{\mathrm{GW}}(c)
	+
	V_{0,m}^{\mathrm{LTC}}(c)
	+
	\Delta V_{0,m}^{\mathrm{SUR}}(c)
	+
	\Delta V_{0,m}^{\mathrm{DB}}(c),
\end{equation}
where
\[
\Delta V_{0,m}^{\mathrm{GW}}(c)
=
V_{0,m}^{\mathrm{GW}}(c)-V_{0,m}^{\mathrm{GW}}(0),
\]
and analogously for the surrender and death/terminal components. Notice that
no LTC payments are made when \(c=0\), so that
\(V_{0,m}^{\mathrm{LTC}}(0)=0\). Hence, the marginal direct LTC contribution
satisfies
\(\Delta V_{0,m}^{\mathrm{LTC}}(c)=V_{0,m}^{\mathrm{LTC}}(c)\) and is therefore
reported in levels in \eqref{eq:dec_delta}.

From a numerical viewpoint, the decomposition is obtained by propagating four value
functions, one for each cash-flow component, through the same hybrid tree--IMEX
recursion used for the total value. The mixed surrender decision is determined by the
total contract value, not by the individual components. Once the optimal mixed action
has been selected, the resulting cash flow is assigned to the appropriate component:
scheduled withdrawals to \(V^{\mathrm{GW}}\), LTC benefits to \(V^{\mathrm{LTC}}\), surrender
payments to \(V^{\mathrm{SUR}}\), and death or terminal payments to \(V^{\mathrm{DB}}\).
This ensures that the decomposition is evaluated under the same exercise rule as the
total contract value.

\begin{table}[H]
	\centering 
	\small
	\setlength{\tabcolsep}{2.6pt}
	{
		\renewcommand{\arraystretch}{0.94}
		\begin{tabular}{lc rr rr rr rr rr}
			\toprule
			\multirow{2}{*}{Model}
			& \multirow{2}{*}{$c$}
			& \multicolumn{2}{c}{$V_0^{\mathrm{GW}}$}
			& \multicolumn{2}{c}{$V_0^{\mathrm{LTC}}$}
			& \multicolumn{2}{c}{$V_0^{\mathrm{SUR}}$}
			& \multicolumn{2}{c}{$V_0^{\mathrm{DB}}$}
			& \multicolumn{2}{c}{$V_0$} \\
			\cmidrule(lr){3-4}
			\cmidrule(lr){5-6}
			\cmidrule(lr){7-8}
			\cmidrule(lr){9-10}
			\cmidrule(lr){11-12}
			&
			& Value & $\Delta$
			& Value & $\Delta$
			& Value & $\Delta$
			& Value & $\Delta$
			& Value & $\Delta$ \\
			\midrule
			
			\textbf{GBM--HW}
			& $0$
			& $25.57$ & -- 
			& $0.00$  & --
			& $61.55$ & --
			& $12.88$ & --
			& $100.00$ & -- \\
			
			& $0.03$
			& $31.88$ & $+6.31$
			& $4.11$  & $+4.11$
			& $45.36$ & $-16.19$
			& $20.76$ & $+7.87$
			& $102.11$ & $+2.11$ \\
			
			& $0.06$
			& $36.15$ & $+10.58$
			& $10.10$ & $+10.10$
			& $30.01$ & $-31.54$
			& $28.96$ & $+16.08$
			& $105.22$ & $+5.22$ \\
			
			\addlinespace[3pt]
			
			\textbf{MJD--HW}
			& $0$
			& $27.01$ & --
			& $0.00$  & --
			& $60.58$ & --
			& $12.41$ & --
			& $100.00$ & -- \\
			
			& $0.03$
			& $31.61$ & $+4.60$
			& $4.05$  & $+4.05$
			& $49.02$ & $-11.55$
			& $17.76$ & $+5.34$
			& $102.44$ & $+2.44$ \\
			
			& $0.06$
			& $34.93$ & $+7.93$
			& $9.59$  & $+9.59$
			& $37.88$ & $-22.70$
			& $23.33$ & $+10.91$
			& $105.73$ & $+5.73$ \\
			
			\addlinespace[3pt]
			
			\textbf{VG--HW}
			& $0$
			& $26.94$ & --
			& $0.00$  & --
			& $60.59$ & --
			& $12.47$ & --
			& $100.00$ & -- \\
			
			& $0.03$
			& $31.64$ & $+4.70$
			& $4.06$  & $+4.06$
			& $48.81$ & $-11.78$
			& $17.92$ & $+5.45$
			& $102.43$ & $+2.43$ \\
			
			& $0.06$
			& $35.01$ & $+8.06$
			& $9.62$  & $+9.62$
			& $37.50$ & $-23.09$
			& $23.58$ & $+11.12$
			& $105.71$ & $+5.71$ \\
			
			\addlinespace[3pt]
			
			\textbf{CGMY--HW}
			& $0$
			& $26.19$ & --
			& $0.00$  & --
			& $61.06$ & --
			& $12.75$ & --
			& $100.00$ & -- \\
			
			& $0.03$
			& $31.74$ & $+5.56$
			& $4.08$  & $+4.08$
			& $46.97$ & $-14.09$
			& $19.45$ & $+6.70$
			& $102.25$ & $+2.25$ \\
			
			& $0.06$
			& $35.59$ & $+9.40$
			& $9.87$  & $+9.87$
			& $33.59$ & $-27.47$
			& $26.38$ & $+13.64$
			& $105.43$ & $+5.43$ \\
			
			\bottomrule
		\end{tabular}
	}
	
	\caption{\small\label{tab:value_decomposition_mixed}Value decomposition under the mixed withdrawal strategy. Values are obtained with IMEX configuration D. 
		For each model, the account-value fee is calibrated in the no-LTC benchmark under the mixed withdrawal strategy and then kept fixed when the LTC payout rate \(c\) is increased. 
		The components \(V_0^{\mathrm{GW}}\), \(V_0^{\mathrm{LTC}}\), \(V_0^{\mathrm{SUR}}\), and \(V_0^{\mathrm{DB}}\) denote, respectively, the discounted values of guaranteed withdrawals, direct LTC payments, surrender payments, and death or terminal payments. 
		For \(c=0.03\) and \(c=0.06\), the columns labelled \(\Delta\) report changes relative to the corresponding \(c=0\) benchmark for the same model.  
	}
	\normalsize
\end{table}

Table~\ref{tab:value_decomposition_mixed} shows that the introduction of LTC
coverage changes not only the level of the contract value, but also its
internal composition. As \(c\) increases, the direct LTC component becomes
positive, while the values of guaranteed withdrawals and death or terminal
payments also increase. At the same time, the surrender component decreases
substantially across all financial specifications. This indicates that LTC
coverage makes continuation more valuable relative to surrender: part of the
value that would otherwise be realized through early termination is shifted
toward health-contingent payments, future guaranteed withdrawals, and residual
terminal benefits.

The net increase in \(V_0\) is therefore much smaller than the discounted value
of direct LTC payments alone. For instance, at \(c=0.03\), the direct LTC
component is about four percentage points of the initial premium across models,
whereas the total contract value increases by only about two to two and a half
percentage points. This difference is explained by the negative contribution of
the surrender component, which partially offsets the positive effects of LTC
payments, guaranteed withdrawals, and death or terminal benefits. Hence, the
marginal value of LTC coverage should be interpreted as an interaction effect
between actuarial benefits and policyholder exercise behaviour, rather than as
a purely additive cash-flow component.

\subsection{Impact of jumps and stochastic rates}
\label{subsec:financial_effects}
\paragraph{A two-way decomposition of financial modelling effects}

All fair fees in this subsection are computed under the mixed withdrawal
strategy. This choice keeps the policyholder exercise rule fixed while allowing
for surrender optionality, and therefore isolates the effect of replacing the
GBM deterministic-rate benchmark with calibrated L\'evy fund dynamics and
Hull--White stochastic rates.

Let \(\alpha_{m,\mathrm{det}}^{\mathrm{mix,fair}}(c)\) and
\(\alpha_{m,\mathrm{HW}}^{\mathrm{mix,fair}}(c)\) denote the fair annual
account-value fees under equity model \(m\), deterministic or Hull--White
interest rates, and the mixed withdrawal strategy. To disentangle the
contribution of stochastic interest rates from that of the equity-return
specification, we use a simple two-factor accounting decomposition. The
benchmark is the GBM specification with deterministic rates. The first factor
replaces deterministic rates with Hull--White rates, while the second replaces
the GBM return distribution with the calibrated L\'evy specification. The
remaining cross-effect is reported as an interaction term.

We decompose the total correction relative to the GBM deterministic-rate
benchmark into three terms:
\[
\Delta_\alpha^{\mathrm{rate}}(c)
=
\alpha_{\mathrm{GBM},\mathrm{HW}}^{\mathrm{mix,fair}}(c)
-
\alpha_{\mathrm{GBM},\mathrm{det}}^{\mathrm{mix,fair}}(c),
\]
\[
\Delta_{\alpha,m}^{\mathrm{dist}}(c)
=
\alpha_{m,\mathrm{det}}^{\mathrm{mix,fair}}(c)
-
\alpha_{\mathrm{GBM},\mathrm{det}}^{\mathrm{mix,fair}}(c),
\]
and
\[
\Delta_{\alpha,m}^{\mathrm{int}}(c)
=
\alpha_{m,\mathrm{HW}}^{\mathrm{mix,fair}}(c)
-
\alpha_{m,\mathrm{det}}^{\mathrm{mix,fair}}(c)
-
\alpha_{\mathrm{GBM},\mathrm{HW}}^{\mathrm{mix,fair}}(c)
+
\alpha_{\mathrm{GBM},\mathrm{det}}^{\mathrm{mix,fair}}(c).
\]
The total correction is therefore
\[
\Delta_{\alpha,m}^{\mathrm{tot}}(c)
=
\Delta_\alpha^{\mathrm{rate}}(c)
+
\Delta_{\alpha,m}^{\mathrm{dist}}(c)
+
\Delta_{\alpha,m}^{\mathrm{int}}(c).
\]
Equivalently,
\[
\Delta_{\alpha,m}^{\mathrm{tot}}(c)
=
\alpha_{m,\mathrm{HW}}^{\mathrm{mix,fair}}(c)
-
\alpha_{\mathrm{GBM},\mathrm{det}}^{\mathrm{mix,fair}}(c).
\]
Thus, the decomposition is an exact algebraic identity, and the interaction
term measures the part of the joint effect that is not obtained by adding the
stochastic-rate and distributional effects separately.

Table~\ref{tab:two_way_financial_decomposition} reports the decomposition of
the fair-fee correction relative to the GBM deterministic-rate benchmark. Panel
A isolates the stochastic-rate effect in the GBM case, while Panel B reports,
for each L\'evy model \(m\), the distributional correction
\(\Delta_{\alpha,m}^{\mathrm{dist}}(c)\), the interaction term
\(\Delta_{\alpha,m}^{\mathrm{int}}(c)\), the total correction
\(\Delta_{\alpha,m}^{\mathrm{tot}}(c)\), and the complete L\'evy--Hull--White
fair fee \(\alpha_{m,\mathrm{HW}}^{\mathrm{mix,fair}}(c)\). The table shows
that both financial dimensions materially affect the fair fee. In particular,
replacing deterministic rates with Hull--White rates increases the GBM fair fee
for all LTC payout rates, with a correction that rises from \(12.34\) bps when
\(c=0\) to \(19.47\) bps and \(26.46\) bps when \(c=0.03\) and \(c=0.06\),
respectively. This indicates that stochastic interest rates have a larger pricing impact as
the LTC payout rate increases, because higher LTC benefits make the contract
more exposed to long-dated cash flows and hence to interest-rate fluctuations.

Panel B shows that the calibrated distributional effect is also positive for all
models and payout rates. The largest corrections are obtained under MJD and VG
dynamics. For example, when \(c=0.06\), replacing the GBM benchmark with MJD or
VG under deterministic rates increases the fair fee by 27.69 bps and 25.33 bps,
respectively. The CGMY specification produces a smaller correction for positive
LTC payout rates and remains closer to the GBM benchmark.

The interaction term is generally smaller than the two main effects and is mostly negative when \(c>0\). 
This means that the impact of the calibrated Lévy distribution and that of stochastic interest rates are close to additive, but their joint effect is slightly lower than the sum of the two effects computed separately. 
Overall, the table confirms that the full Lévy--Hull--White fair fee cannot be attributed to a single modelling feature. 
It reflects the combined contribution of calibrated fund-return distributions, stochastic interest rates, and their interaction.

Since the \citet{bacinello2016} parameter sets are not moment-matched across models, 
\(\Delta_{\alpha,m}^{\mathrm{dist}}\) should not be interpreted as a pure jump premium. 
Rather, it captures the combined effect of jumps, volatility, skewness, kurtosis, and tail behaviour implied by the option calibration.
\begin{table}[H]
	\centering
	\small
	\setlength{\tabcolsep}{4pt}
	\renewcommand{\arraystretch}{0.94}
	
	\begin{tabular*}{0.6\textwidth}{@{\extracolsep{\fill}}c rrr@{}}
		\toprule
		\multicolumn{4}{@{}l}{\textit{Panel A: \textbf{GBM} benchmark and stochastic-rate effect}} \\
		\midrule
		\(c\)
		& \(\alpha_{\mathrm{GBM},\mathrm{det}}^{\mathrm{mix,fair}}\)
		& \(\alpha_{\mathrm{GBM},\mathrm{HW}}^{\mathrm{mix,fair}}\)
		& \(\Delta_\alpha^{\mathrm{rate}}\) \\
		\midrule
		$0$    & $-10.18$ & $2.16$  & $+12.34$ \\
		$0.03$ & $5.54$   & $25.01$ & $+19.47$ \\
		$0.06$ & $32.19$  & $58.65$ & $+26.46$ \\
		\bottomrule
	\end{tabular*}
	
	\vspace{2mm}
	
	\begin{tabular*}{0.6\textwidth}{@{\extracolsep{\fill}}l c rrrr@{}}
		\toprule
		\multicolumn{6}{@{}l}{\textit{Panel B: L\'evy distributional and interaction effects}} \\
		\midrule
		L\'evy Model 
		& \(c\)
		& \(\Delta_{\alpha,m}^{\mathrm{dist}}\)
		& \(\Delta_{\alpha,m}^{\mathrm{int}}\)
		& \(\Delta_{\alpha,m}^{\mathrm{tot}}\)
		& \(\alpha_{m,\mathrm{HW}}^{\mathrm{mix,fair}}\) \\
		\midrule
		
		\textbf{MJD}
		& $0$    & $+15.09$ & $+0.13$  & $+27.56$ & $17.38$ \\
		& $0.03$ & $+21.78$ & $-1.36$  & $+39.89$ & $45.43$ \\
		& $0.06$ & $+27.69$ & $-2.54$  & $+51.61$ & $83.80$ \\
		
		\addlinespace[3pt]
		
		\textbf{VG}
		& $0$    & $+13.28$ & $+0.28$  & $+25.90$ & $15.72$ \\
		& $0.03$ & $+19.62$ & $-1.17$  & $+37.92$ & $43.46$ \\
		& $0.06$ & $+25.33$ & $-2.36$  & $+49.43$ & $81.62$ \\
		
		\addlinespace[3pt]
		
		\textbf{CGMY}
		& $0$    & $+15.75$ & $-10.70$ & $+17.39$ & $7.21$ \\
		& $0.03$ & $+7.41$  & $-0.31$  & $+26.57$ & $32.11$ \\
		& $0.06$ & $+9.99$  & $-0.96$  & $+35.49$ & $67.68$ \\
		
		\bottomrule
	\end{tabular*}
	
	\caption{\small \label{tab:two_way_financial_decomposition}
		Two-way decomposition of fair fees under calibrated L\'evy distributions
		and stochastic interest rates. Fair fees are reported in basis points and
		are computed under the mixed withdrawal strategy. The benchmark model is
		the GBM specification with deterministic interest rates. The last column
		of Panel B reports \(\alpha_{m,\mathrm{HW}}^{\mathrm{mix,fair}}(c)\),
		the fair fee under the complete L\'evy--Hull--White specification.
	}
\end{table}

\paragraph{Sensitivity analysis under the mixed strategy}
\label{subsubsec:sensitivity_mixed}

We next investigate the sensitivity of the mixed-strategy fair annual fee
\(\alpha_{\mathrm{MJD\text{--}HW}}^{\mathrm{mix,fair}}(c)\) in the MJD--HW
model. We focus on the MJD specification because its
jump parameters have a direct interpretation in terms of jump intensity, mean
jump size, and jump-size volatility. 
For brevity, throughout this sensitivity analysis, we write
\(\alpha^{\mathrm{fair}}\) for
\(\alpha_{\mathrm{MJD\text{--}HW}}^{\mathrm{mix,fair}}(c)\).

The analysis is performed by varying one parameter at a time, while keeping all
remaining parameters fixed at their benchmark values. For each parameter value,
the fair fee is recomputed by imposing the at-inception fairness condition.
Fair fees are reported in basis points.

\begin{figure}[t]
	\centering
	
	\begin{subfigure}{0.49\textwidth}
		\centering
		\includegraphics[width=\linewidth]{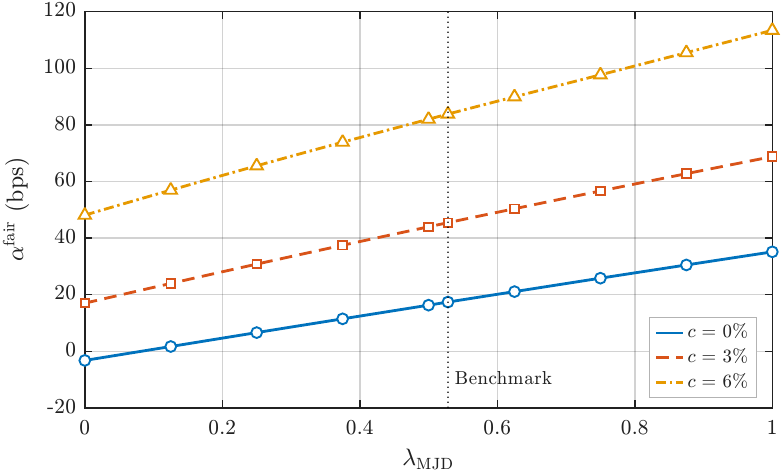}
		\caption{Sensitivity with respect to $\lambda_{\mathrm{MJD}}$}
		\label{fig:mjd_sensitivity_lambda}
	\end{subfigure}
	\hfill
	\begin{subfigure}{0.49\textwidth}
		\centering
		\includegraphics[width=\linewidth]{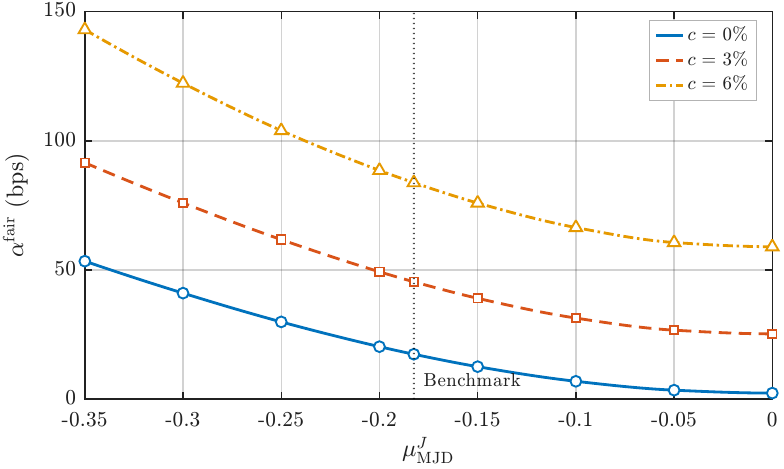}
		\caption{Sensitivity with respect to $\mu^J_{\mathrm{MJD}}$}
		\label{fig:mjd_sensitivity_mu}
	\end{subfigure}
	
	\vspace{2mm}
	
	\begin{subfigure}{0.49\textwidth}
		\centering
		\includegraphics[width=\linewidth]{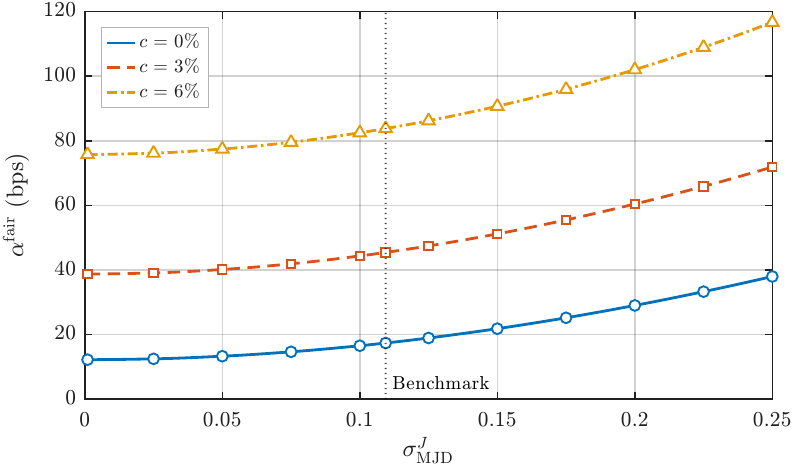}
		\caption{Sensitivity with respect to $\sigma^J_{\mathrm{MJD}}$}
		\label{fig:mjd_sensitivity_sigmaj}
	\end{subfigure}
	
	\caption{\small Sensitivity of the fair annual fee $\alpha^{\mathrm{fair}}$
		to the MJD parameters under the mixed strategy. Fair
		fees are reported in basis points for different values of the LTC payout rate
		 $c$. The vertical dotted line denotes the benchmark parameter
		value.}
	\label{fig:mjd_sensitivity_mixed}
\end{figure}

Figure~\ref{fig:mjd_sensitivity_mixed} shows the impact of the MJD parameters on the fair fee. The fair fee is increasing with
respect to the jump intensity $\lambda_{\mathrm{MJD}}$ and the jump-size
volatility $\sigma^J_{\mathrm{MJD}}$. More frequent jumps and more volatile jump
sizes increase the value of the embedded guarantees, and therefore require a
larger fair fee. Conversely, $\alpha^{\mathrm{fair}}$ generally decreases as
the average jump size $\mu^J_{\mathrm{MJD}}$ becomes less negative over the
relevant benchmark region. This behaviour is consistent with the fact that,
when jumps are less severe on average, the downside risk borne by the insurer
is reduced. 
The three curves corresponding to $c=0\%$, $c=3\%$, and $c=6\%$ display the same
qualitative pattern. Increasing the LTC payout rate $c$ shifts the fair fee
upward, but it does not alter the qualitative dependence of the fair fee on the
jump parameters. This suggests that the effect of jump risk and the effect of
the LTC payout rate are complementary: larger values of $c$ increase the
overall level of the fair fee, while the MJD parameters mainly determine how
sensitive the contract value is to market downside risk.

\begin{figure}[t]
	\centering
	
	\begin{subfigure}{0.49\textwidth}
		\centering
		\includegraphics[width=\linewidth]{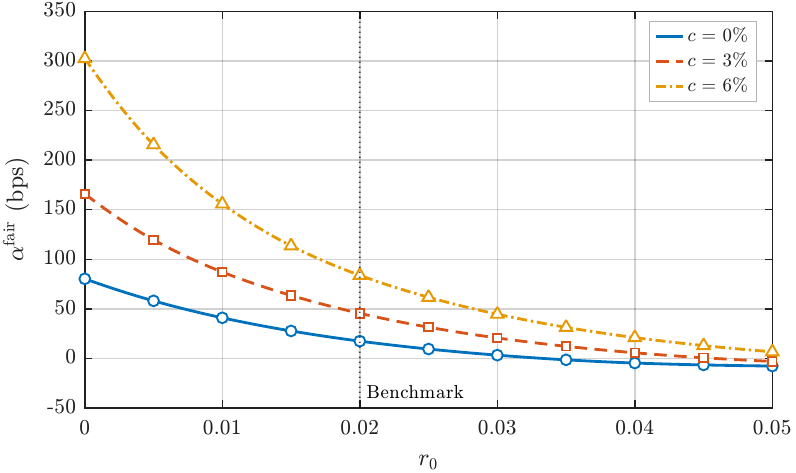}
		\caption{Sensitivity with respect to $r_0$}
		\label{fig:hw_sensitivity_r0}
	\end{subfigure}
	\hfill
	\begin{subfigure}{0.49\textwidth}
		\centering
		\includegraphics[width=\linewidth]{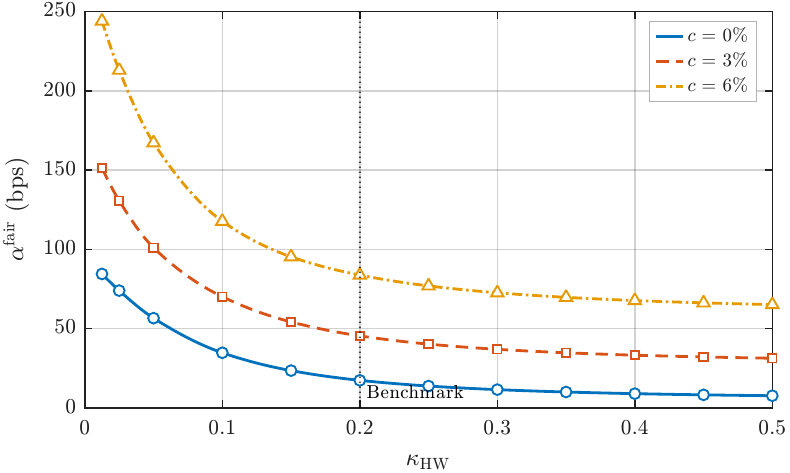}
		\caption{Sensitivity with respect to $\kappa_{\mathrm{HW}}$}
		\label{fig:hw_sensitivity_kappa}
	\end{subfigure}
	
	\vspace{2mm}
	
	\begin{subfigure}{0.49\textwidth}
		\centering
		\includegraphics[width=\linewidth]{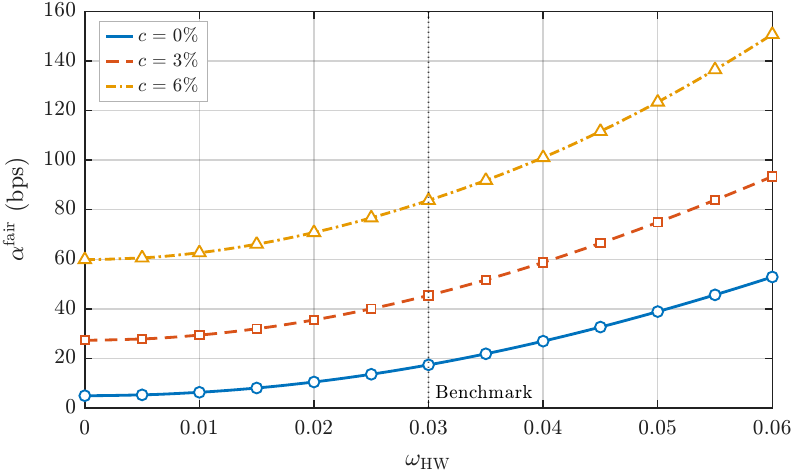}
		\caption{Sensitivity with respect to $\omega_{\mathrm{HW}}$}
		\label{fig:hw_sensitivity_omega}
	\end{subfigure}
	
	\caption{\small\label{fig:hw_sensitivity_mixed}Sensitivity of the fair annual fee $\alpha^{\mathrm{fair}}$
		to the Hull--White parameters under the mixed strategy. Fair fees are
		reported in basis points for different values of the LTC payout rate
		$c$. The vertical dotted line denotes the benchmark parameter value.}
	
\end{figure}

Figure~\ref{fig:hw_sensitivity_mixed} reports the corresponding sensitivities
with respect to the Hull--White parameters. The fair fee decreases as the
initial short rate $r_0$ increases. This effect is particularly pronounced,
reflecting the relevance of the interest-rate environment for a long-dated
insurance contract. Higher initial rates increase the discounting of future
liabilities and therefore reduce the fee required to make the contract fair at
inception.

The fair fee also decreases with the mean-reversion speed
$\kappa_{\mathrm{HW}}$. The effect is stronger for low values of
$\kappa_{\mathrm{HW}}$ and becomes progressively flatter as the mean-reversion
speed increases. Economically, stronger mean reversion limits the persistence
of deviations in the short rate, reducing the impact of adverse interest-rate
scenarios on the value of the guarantees.

By contrast, the fair fee increases with the short-rate volatility
$\omega_{\mathrm{HW}}$. A larger short-rate volatility increases the value of
the guarantees by giving more weight to adverse low-rate scenarios, in which
future liabilities are discounted less heavily. As in the MJD sensitivity
analysis, the curves associated with larger values of $c$ lie above those
obtained for smaller values of $c$, while preserving the same qualitative
dependence on the underlying model parameters.

Overall, the sensitivity results confirm that the fair fee under the mixed
strategy is affected by both sources of risk. The MJD parameters mainly control
the exposure to equity downside jump risk, while the Hull--White parameters
affect the valuation through the stochastic discounting channel. Across all
experiments, the LTC payout rate $c$ has a level effect on the fair fee:
higher payout rates require higher fees, but do not materially change the
qualitative comparative statics with respect to the financial model parameters.

\paragraph{Effect of jump intensity on the surrender boundary}
\label{subsubsec:mjd_lambda_surrender_boundary}

We finally investigate how jump intensity affects the optimal surrender
decision under the mixed strategy. The purpose of this test is not to provide a
full analysis of optimal exercise regions, but rather to assess whether higher
jump intensity changes the policyholder's surrender incentives within the
MJD--HW framework.

To isolate the effect of jump intensity on the exercise decision, the test is
performed under the mixed strategy, with \(c=0.03\), at year \(n=10\), and for a
policyholder in the healthy state \(M_{10}=1\). The account-value fee
\(\alpha\) is kept fixed at the corresponding benchmark mixed-strategy fair
level,
\(\alpha_{\mathrm{MJD\text{--}HW}}^{\mathrm{mix,fair}}(c=0.03)=45.43\)
bps, computed with configuration D. Hence, \(\alpha\) is not recalibrated when
\(\lambda_{\mathrm{MJD}}\) changes, so that changes in the surrender boundary
reflect the effect of the jump-intensity parameter itself rather than the
indirect effect of a different fair-fee calibration.

Let \(a_n^\star(r)\) denote the critical post-fee and post-LTC account value
above which surrender is optimal, conditional on the short rate \(r\) at
anniversary \(n\). Using the notation of the annual contract operator in
Section~\ref{subsec:annual_operator}, the boundary at the mixed-strategy
decision stage is characterised by
\[
a_n^\star(r)
=
\inf\left\{
a^{(2)}\geq 0:
Y_n^{(2)}
\geq
Y_n^{(1)}
+
\widetilde V_n\!\left(a_n^{+,1},b_n^{+,1},m,r\right)
\right\}.
\]
Here the action-dependent quantities are evaluated at the decision-stage state
\((a^{(2)},b^{(2)},m,r)\). The boundary is therefore computed after fees and
LTC payments, and immediately before the withdrawal/surrender decision. For
account values above \(a_n^\star(r)\), the surrender value exceeds the value of
taking the scheduled guaranteed withdrawal and continuing the contract, while
for lower account values continuation is optimal.

\begin{figure}[t]
	\centering
	\includegraphics[width=0.65\textwidth]{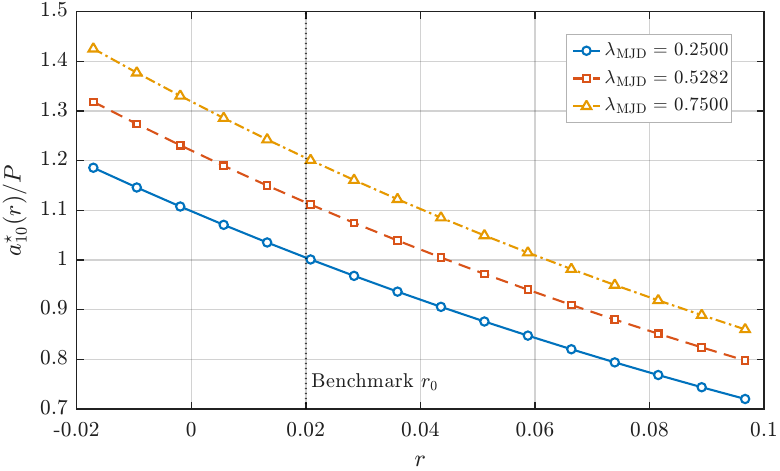}
\caption{\small \label{fig:mjd_lambda_surrender_boundary}
	Effect of MJD jump intensity on the surrender boundary under the mixed
	strategy. The figure reports the critical post-fee and post-LTC account-value
	ratio \(a_{10}^{\star}(r)/P\) as a function of the short rate \(r\), for
	\(c=0.03\) and \(M_{10}=1\). Contract parameters, including \(\alpha\), are kept
	fixed at their benchmark MJD--HW values.}
\end{figure}

Figure~\ref{fig:mjd_lambda_surrender_boundary} shows that increasing the jump
intensity shifts the surrender boundary upward. Hence, when jumps become more
frequent, the policyholder requires a larger account value before surrender
becomes optimal. This behaviour is consistent with the insurance role of the
embedded guarantee: under stronger jump risk, continuation becomes relatively
more valuable because surrendering the contract implies giving up protection
against future account depletion in adverse jump scenarios.

The figure also shows that the surrender boundary is decreasing in the short
rate. Higher short rates reduce the present value of future guarantee payments
and make continuation less attractive relative to immediate surrender. Therefore,
surrender becomes optimal at lower account-value levels when the short rate is
higher. Overall, the test confirms that jump risk affects not only the fair fee,
but also the policyholder's optimal exercise incentives. In particular, stronger
jump intensity increases the continuation value of the contract and reduces the
region in which surrender is optimal.

\FloatBarrier

\bigskip
The numerical evidence indicates that the L\'evy--Hull--White specification
generates non-negligible corrections to the valuation of GLWB--LTC contracts
relative to simpler benchmarks. The hybrid tree--IMEX method remains stable on
long horizons and delivers values consistent with Monte Carlo, while being well
suited for repeated pricing and comparative statics. From an actuarial
perspective, the interaction between jump risk, stochastic discounting, and
LTC-contingent withdrawals materially affects fair-fee determination and should
not be ignored in product design.

\FloatBarrier
 
\section{Conclusion}
\label{sec:conclusion}

In this paper, we have considered the valuation of GLWB--LTC contracts under
exponential L\'evy dynamics for the reference fund and Hull--White dynamics for
the short rate. The proposed framework combines a multi-state disability
process, annual GLWB--LTC contract mechanics, and a hybrid tree--IMEX method.
In doing so, the analysis brings together health-contingent GLWB--LTC valuation
and L\'evy--Hull--White numerical methods for long-term variable annuity
guarantees.

Within the numerical setting considered, the hybrid method exhibits stable
behaviour across grid refinements and produces prices consistent with Monte
Carlo benchmarks in the static case. The numerical experiments also indicate that replacing simpler financial
benchmarks with calibrated L\'evy equity dynamics and Hull--White stochastic
rates can have a non-negligible impact on fair fees.
 The value decomposition suggests that the LTC rider
does not behave as a purely additive cash-flow component: in the specifications
analysed, LTC payments interact with account depletion, surrender incentives,
guaranteed withdrawals, and death or terminal payments. The sensitivity and
surrender-boundary analyses further illustrate that jump-risk parameters may
affect not only the level of the fair fee, but also the policyholder's exercise
incentives.

Overall, the results provide numerical evidence that financial tail risk,
stochastic discounting, and health-contingent withdrawals can interact in
economically relevant ways when valuing long-term insurance guarantees with LTC
benefits. At the same time, the reported fee levels and comparative statics are conditional on the adopted
financial, actuarial, and behavioural assumptions and exclude explicit loadings
for biometric risk, expenses, capital costs, or profit margins. They should
therefore be interpreted as evidence on model sensitivity and contract
interactions, rather than as universal pricing levels.

\paragraph{Competing interests}
The author declares no competing interests.

\paragraph{Funding statement}
This research received no specific grant from any funding agency, commercial or
not-for-profit sectors.

\paragraph{Data availability statement}
The numerical experiments are based on the parameter values reported in the
paper and on the  health-transition probabilities described in
 \citet{pritchard2006}. The input files used to generate the
health-transition matrices and the numerical results are available from the
author upon request.

\paragraph{Code availability statement}
The MATLAB code used to generate the numerical results is available from the
author upon request.
 
\bibliographystyle{apalike}
\bibliography{paper_glwb_ltc_levy_hw}

\appendix

\section{Log-return moment calculation with Hull--White interest rates}
\label{app:hw_moment_calculation}

This appendix summarises the moment calculation used in
Table~\ref{tab:bacinello_hw_rn_moments}. Throughout the appendix, \(T>0\) denotes a generic time horizon over which the
log-return moments are computed. Let
\[
I_T=\int_0^T r_s\,ds
\]
be the integrated short rate. Under the exponential L\'evy--Hull--White model,
the log-return over the horizon \(T\) is
\[
Y_T
=
\log\frac{S_T}{S_0}
=
I_T-qT+X_T-K_X(1)T,
\]
where \(q\) is the dividend yield, \(X_T\) is the L\'evy component, and
\(K_X(1)\) is the exponential martingale compensator, with
\[
\mathbb{E}\!\left[e^{zX_T}\right]=\exp\{TK_X(z)\}.
\]

The short rate follows the one-factor Hull--White model in \eqref{eq:sde_HW}.
In the numerical experiments reported in the paper, the initial zero-coupon curve
is flat, so that \(P(0,T)=e^{-r_0T}\). Under Hull--White dynamics, \(I_T\) is
normally distributed and
\[
\operatorname{Var}(I_T)
=
\frac{\omega^2}{\kappa^2}
\left[
T
-\frac{2(1-e^{-\kappa T})}{\kappa}
+\frac{1-e^{-2\kappa T}}{2\kappa}
\right].
\]
Since \(I_T\) is Gaussian,
\[
P(0,T)
=
\mathbb{E}\!\left[e^{-I_T}\right]
=
\exp\left(
-\mathbb{E}[I_T]+\frac{1}{2}\operatorname{Var}(I_T)
\right).
\]
Therefore, in the flat-curve case used in the numerical section,
\[
\mathbb{E}[I_T]
=
r_0T+\frac{1}{2}\operatorname{Var}(I_T).
\]
For a non-flat initial curve, the same calculation applies by replacing
\(e^{-r_0T}\) with the observed initial discount factor \(P(0,T)\), namely
\[
\mathbb{E}[I_T]
=
-\log P(0,T)+\frac{1}{2}\operatorname{Var}(I_T).
\]

Assuming independence between the Hull--White short-rate process and the L\'evy
equity component, cumulants are additive. If \(c_n^X\) denotes the \(n\)-th
cumulant of \(X_T\), then the cumulants of \(Y_T\) are
\[
c_1^Y
=
\mathbb{E}[I_T]-qT+c_1^X-K_X(1)T,
\qquad
c_2^Y
=
\operatorname{Var}(I_T)+c_2^X,
\]
and, because the Hull--White contribution is Gaussian,
\[
c_3^Y=c_3^X,
\qquad
c_4^Y=c_4^X.
\]
The standardised skewness and kurtosis are then
\[
\operatorname{Skewness}(Y_T)
=
\frac{c_3^Y}{(c_2^Y)^{3/2}},
\qquad
\operatorname{Kurtosis}(Y_T)
=
3+\frac{c_4^Y}{(c_2^Y)^2}.
\]

Thus, the Hull--White component affects the first two cumulants directly, and
also affects standardised skewness and kurtosis through the total variance. A
non-zero dependence structure between stochastic interest rates and general
L\'evy equity dynamics would require specifying a joint model and is not used in
the moment calculation reported here.

\end{document}